\pdfoutput=1  

\documentclass[11pt]{article}
\usepackage[margin=.8in,left=.8in]{geometry}
\usepackage{amsmath}
\usepackage{amsfonts}
\usepackage{amssymb}
\usepackage{amsthm}
\usepackage{mathtools}
\usepackage{subcaption}
\usepackage{graphicx}
\usepackage{color}
\usepackage{xcolor}
\usepackage{xfrac}
\usepackage{stmaryrd}

\usepackage[safe]{tipa} 
\usepackage[utf8]{inputenc}
\usepackage{rotating}
\usepackage{hyperref}
\usepackage[all]{xy}

\usepackage{mathtools}
\usepackage{rotating}

\usepackage{stmaryrd}
\usepackage{multirow}

\usepackage{lmodern} 

\usepackage{tikz}
\usetikzlibrary{decorations.pathmorphing}
\usepackage{tikz-cd}
\usetikzlibrary{calc}
\usetikzlibrary{matrix}
\usetikzlibrary{positioning,arrows,shapes}
\usetikzlibrary{intersections,shapes.arrows, calc}
\usetikzlibrary{patterns}
\usetikzlibrary{mindmap} 
\usetikzlibrary[mindmap] 

\usepackage{lipsum}
\usepackage{adjustbox}

\definecolor{redi}{RGB}{255,38,0}
\definecolor{redii}{RGB}{200,50,30}
\definecolor{yellowi}{RGB}{255,251,0}
\definecolor{bluei}{RGB}{0,150,255}
\definecolor{blueii}{RGB}{135,247,210}
\definecolor{blueiii}{RGB}{91,205,250}
\definecolor{blueiv}{RGB}{115,244,253}
\definecolor{bluev}{RGB}{1,58,215}
\definecolor{orangei}{RGB}{220,160, 20}
\definecolor{orangeii}{RGB}{240,90, 10}
\definecolor{yellowii}{RGB}{222,247,100}
\definecolor{greeni}{RGB}{85,102,0}
\definecolor{greenii}{RGB}{20,140,10}
\definecolor{navy}{RGB}{17, 10, 102}
\definecolor{brown}{RGB}{60, 40, 0}
\definecolor{oxford}{RGB}{0, 0, 100}

\definecolor{plum}{rgb}{0.36078, 0.20784, 0.4}
\definecolor{chameleon}{rgb}{0.30588, 0.60392, 0.023529}
\definecolor{cornflower}{rgb}{0.12549, 0.29020, 0.52941}
\definecolor{scarlet}{rgb}{0.8, 0, 0}
\definecolor{brick}{rgb}{0.64314, 0, 0}
\definecolor{sunrise}{rgb}{0.80784, 0.36078, 0}
\definecolor{lightblue}{rgb}{0.15,0.35,0.75}
\definecolor{carolina}{RGB}{153, 186, 221}
\definecolor{darkblue}{rgb}{0.05,0.25,0.65}

\usepackage{multirow}

\usepackage{stmaryrd}    

\usepackage{enumerate} 

\usepackage{helvet}   

\usepackage{mathptmx}
\usepackage{amsmath}
\usepackage{graphicx}

\usepackage{floatflt}  
\usepackage{array}     
\newcolumntype{L}[1]{>{\raggedright\let\newline\\\arraybackslash\hspace{0pt}}m{#1}}
\newcolumntype{C}[1]{>{\centering\let\newline\\\arraybackslash\hspace{0pt}}m{#1}}
\newcolumntype{R}[1]{>{\raggedleft\let\newline\\\arraybackslash\hspace{0pt}}m{#1}}

\newcommand{\gt}{>}

\makeatletter
\newcommand{\raisemath}[1]{\mathpalette{\raisem@th{#1}}}
\newcommand{\raisem@th}[3]{\raisebox{#1}{$#2#3$}}
\makeatother

\usepackage{tabularx}   

\usepackage[new]{old-arrows}   

\newdir{> }{{}*!/10pt/@{>}}

\DeclareRobustCommand{\rchi}{{\mathpalette\irchi\relax}}
\newcommand{\irchi}[2]{\raisebox{\depth}{$#1\chi$}} 

\makeatletter
\newif\if@sup
\newtoks\@sups
\def\append@sup#1{\edef\act{\noexpand\@sups={\the\@sups #1}}\act}%
\def\reset@sup{\@supfalse\@sups={}}%
\def\mk@scripts#1#2{\if #2/ \if@sup ^{\the\@sups}\fi \else%
  \ifx #1_ \if@sup ^{\the\@sups}\reset@sup \fi {}_{#2}%
  \else \append@sup#2 \@suptrue \fi%
  \expandafter\mk@scripts\fi}
\def\tensor#1#2{\reset@sup#1\mk@scripts#2_/}
\def\multiscripts#1#2#3{\reset@sup{}\mk@scripts#1_/#2%
  \reset@sup\mk@scripts#3_/}
\makeatother

\makeatletter
\newbox\slashbox \setbox\slashbox=\hbox{$/$}
\def\itex@pslash#1{\setbox\@tempboxa=\hbox{$#1$}
  \@tempdima=0.5\wd\slashbox \advance\@tempdima 0.5\wd\@tempboxa
  \copy\slashbox \kern-\@tempdima \box\@tempboxa}
\def\slash{\protect\itex@pslash}
\makeatother

\def\clap#1{\hbox to 0pt{\hss#1\hss}}
\def\mathllap{\mathpalette\mathllapinternal}
\def\mathrlap{\mathpalette\mathrlapinternal}
\def\mathclap{\mathpalette\mathclapinternal}
\def\mathllapinternal#1#2{\llap{$\mathsurround=0pt#1{#2}$}}
\def\mathrlapinternal#1#2{\rlap{$\mathsurround=0pt#1{#2}$}}
\def\mathclapinternal#1#2{\clap{$\mathsurround=0pt#1{#2}$}}

\let\oldroot\root
\def\root#1#2{\oldroot #1 \of{#2}}
\renewcommand{\sqrt}[2][]{\oldroot #1 \of{#2}}

\DeclareSymbolFont{symbolsC}{U}{txsyc}{m}{n}
\SetSymbolFont{symbolsC}{bold}{U}{txsyc}{bx}{n}
\DeclareFontSubstitution{U}{txsyc}{m}{n}

\DeclareSymbolFont{stmry}{U}{stmry}{m}{n}
\SetSymbolFont{stmry}{bold}{U}{stmry}{b}{n}

\DeclareFontFamily{OMX}{MnSymbolE}{}
\DeclareSymbolFont{mnomx}{OMX}{MnSymbolE}{m}{n}
\SetSymbolFont{mnomx}{bold}{OMX}{MnSymbolE}{b}{n}
\DeclareFontShape{OMX}{MnSymbolE}{m}{n}{
    <-6>  MnSymbolE5
   <6-7>  MnSymbolE6
   <7-8>  MnSymbolE7
   <8-9>  MnSymbolE8
   <9-10> MnSymbolE9
  <10-12> MnSymbolE10
  <12->   MnSymbolE12}{}

\makeatletter
\def\Decl@Mn@Delim#1#2#3#4{%
  \if\relax\noexpand#1%
    \let#1\undefined
  \fi
  \DeclareMathDelimiter{#1}{#2}{#3}{#4}{#3}{#4}}
\def\Decl@Mn@Open#1#2#3{\Decl@Mn@Delim{#1}{\mathopen}{#2}{#3}}
\def\Decl@Mn@Close#1#2#3{\Decl@Mn@Delim{#1}{\mathclose}{#2}{#3}}
\Decl@Mn@Open{\llangle}{mnomx}{'164}
\Decl@Mn@Close{\rrangle}{mnomx}{'171}
\Decl@Mn@Open{\lmoustache}{mnomx}{'245}
\Decl@Mn@Close{\rmoustache}{mnomx}{'244}
\makeatother

\makeatletter
\DeclareRobustCommand\widecheck[1]{{\mathpalette\@widecheck{#1}}}
\def\@widecheck#1#2{%
    \setbox\z@\hbox{\m@th$#1#2$}%
    \setbox\tw@\hbox{\m@th$#1%
       \widehat{%
          \vrule\@width\z@\@height\ht\z@
          \vrule\@height\z@\@width\wd\z@}$}%
    \dp\tw@-\ht\z@
    \@tempdima\ht\z@ \advance\@tempdima2\ht\tw@ \divide\@tempdima\thr@@
    \setbox\tw@\hbox{%
       \raise\@tempdima\hbox{\scalebox{1}[-1]{\lower\@tempdima\box
\tw@}}}%
    {\ooalign{\box\tw@ \cr \box\z@}}}
\makeatother

\makeatletter
\def\udots{\mathinner{\mkern2mu\raise\p@\hbox{.}
\mkern2mu\raise4\p@\hbox{.}\mkern1mu
\raise7\p@\vbox{\kern7\p@\hbox{.}}\mkern1mu}}
\makeatother

\def\1{{\bf 1}}

\def\<{\langle}
\def\>{\rangle}

\renewcommand{\(}{\begin{equation}}
\renewcommand{\)}{\end{equation}}
\newcommand{\bea}{\begin{eqnarray*}}
\newcommand{\eea}{\end{eqnarray*}}

\usepackage{cleveref}

\crefformat{section}{\S#2#1#3} 
\crefformat{subsection}{\S#2#1#3}
\crefformat{subsubsection}{\S#2#1#3}

\theoremstyle{italics}
\newtheorem{theorem}{Theorem}

\theoremstyle{definition}

\newtheorem{note[theorem]}{Note}

\usepackage{amsfonts}
\usepackage{colortbl}

\definecolor{darkblue}{rgb}{0.05,0.25,0.65}
\definecolor{darkgreen}{rgb}{0.00,0.85,0.1}
\definecolor{plum}{rgb}{0.36078, 0.20784, 0.4}

\usepackage{hyperref}

\usepackage{amsfonts}

\begin{document}

\title{Twisted Cohomotopy implies twisted String structure on M5-branes}

\author{
  Domenico Fiorenza, \;
  Hisham Sati, \;
  Urs Schreiber
}

\maketitle

\vspace{-.3cm}

\begin{abstract}
  We show that charge-quantization of the M-theory C-field
  in J-twisted Cohomotopy
  implies emergence of a higher
  $\mathrm{Sp}(1)$-gauge field on single
  heterotic M5-branes, which exhibits worldvolume $\mathrm{String}^{c_2}$-structure.
\end{abstract}

Towards an understanding of the elusive mathematical formulation of
M-theory \cite{Duff96}\cite{Duff99}\cite{BeckerBeckerSchwarz06}
the following assumption has been proposed
\cite{Sati13}\cite{FSS19b}\cite{FSS19c}\cite{SS19a}\cite{SS19c}\cite{SS21}:

\vspace{-.1cm}

\begin{center}
\hypertarget{HypothesisH}{}
\fbox{
  \begin{minipage}[l]{15.5cm}
  {\bf Hypothesis H:}
  \it
  The M-Theory C-field is charge-quantized in J-twisted
  Cohomotopy theory, i.e. the
  C-field flux densities $G_4$, $G_7$
  are in the image of the
  twisted/equivariant cohomotopical character map
  (from \cite[\S 5.3]{FSS20b}\cite[Thm. 1.1]{SS20c}).
  \end{minipage}
}
\end{center}

\vspace{-.1cm}

\noindent Assuming this, fields of the M5-brane sigma-model
\cite{FSS19d}
are found to be encoded by classifying maps shown in
{\color{greenii}green}
in the following homotopy-commutative diagram
of topological spaces \cite{FSS19c} (based on \cite{FSS19b}):

\vspace{2cm}

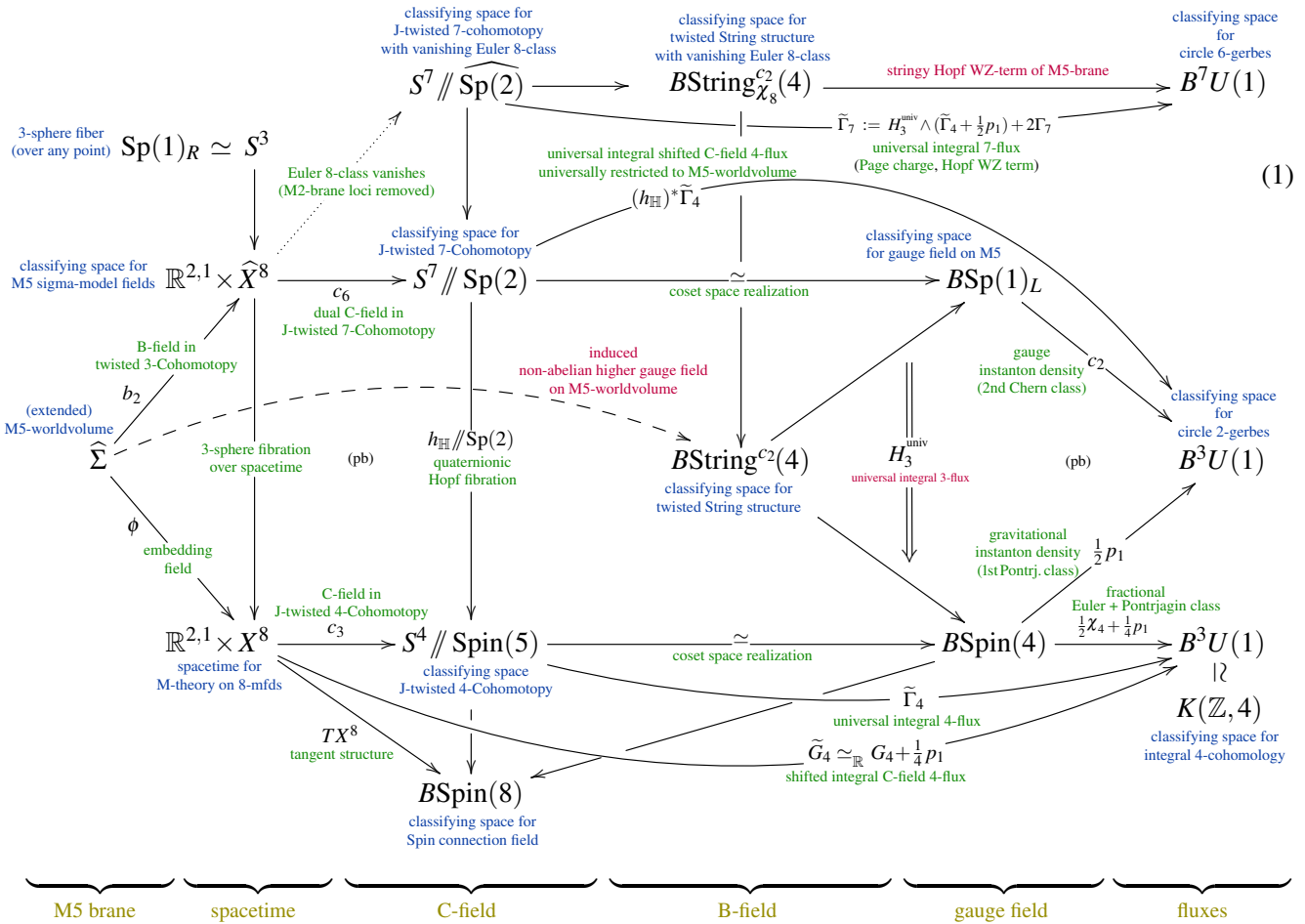
\begin{equation}
  \label{TheFactorization}
\end{equation}

\vspace{-3cm}

 \hspace{-.9cm}
 \hspace{.2cm}
 \scalebox{1}{
 \begin{tikzpicture}

  \draw (-7.75,-5.5) node
   {
     $
       \underset{
         \scalebox{.75}{
           \color{olive}
           \begin{tabular}{c}
             M5 brane
           \end{tabular}
         }
       }{
         \underbrace{
           \phantom{------}
         }
       }
     $
   };
  \draw (-8+2.4,-5.5) node
   {
     $
       \underset{
         \scalebox{.75}{
           \color{olive}
           \begin{tabular}{c}
             spacetime
           \end{tabular}
         }
       }{
         \underbrace{
           \phantom{------}
         }
       }
     $
   };
  \draw (-8+2.4+2.9,-5.5) node
   {
     $
       \underset{
         \scalebox{.75}{
           \color{olive}
           \begin{tabular}{c}
             $\;\;\;\;\;\;\;$C-field
           \end{tabular}
         }
       }{
         \underbrace{
           \phantom{----------}
         }
       }
     $
   };
  \draw (-8+9.1,-5.5) node
   {
     $
       \underset{
         \scalebox{.75}{
           \color{olive}
           \begin{tabular}{c}
             $\;\;\;\;\;\;\;\;\;$B-field
           \end{tabular}
         }
       }{
         \underbrace{
           \phantom{-----------}
         }
       }
     $
   };
  \draw (-8+12.55,-5.5) node
   {
     $
       \underset{
         \scalebox{.75}{
           \color{olive}
           \begin{tabular}{c}
             $\;\;\;$gauge field
           \end{tabular}
         }
       }{
         \underbrace{
           \phantom{--------}
         }
       }
     $
   };
  \draw (-8+15.25,-5.5) node
   {
     $
       \underset{
         \scalebox{.75}{
           \color{olive}
           \begin{tabular}{c}
             $\;\;\;\;\;$fluxes
           \end{tabular}
         }
       }{
         \underbrace{
           \phantom{-------}
         }
       }
     $
   };

 \draw (2.3,4.95) node
 {
   \xymatrix@C=38pt{
     \ar@/_.8pc/[rrrrrr]|>>>>>>>>>>{
       \overset{
         \mbox{
           \tiny
           \begin{tabular}{c}
             \phantom{a}
             \\
             \phantom{a}
           \end{tabular}
         }
       }{
       \underset{
         \mbox{
           \tiny
           \begin{tabular}{c}
             \color{greenii}
             universal integral 7-flux
             \\
             ({\color{greenii}Page charge}, {\color{greenii}Hopf WZ term})
           \end{tabular}
         }
       }{
         \scalebox{.6}{
         $
         \widetilde \Gamma_7
         \;:=\;
         H^{{}^{\mathrm{univ}}}_3 \!
         \wedge
         (\widetilde \Gamma_4 + \tfrac{1}{2}p_1)
         +
         2\Gamma_7
         $
         }
       }
       }
     }
     &&&&&&
   }
 };

 \draw[draw=white, fill=white] (0.975,4.95) circle (.07);

  \draw (0,0) node
  {
  \raisebox{68pt}{
  \xymatrix@C=45pt{
    &
    \mathllap{
      \mathllap{
        \overset{
          \mbox{
            \tiny
            \color{darkblue}
            3-sphere fiber
          }
        }{
        \mbox{
          \tiny
          \color{darkblue}
          (over any point)
        }
        }
        \;
        \mathrm{Sp}(1)_R
        \,\simeq\;
      }
    }
    \;
    S^3
    \mathrlap{
    }
    \ar[d]
    \\
    &
    \mathllap{
      \overset{
        \mbox{
          \tiny
          \color{darkblue}
          classifying space for
        }
      }
      {
        \mbox{
          \tiny
          \color{darkblue}
          M5 sigma-model fields
        }
      }
      \;
      \mathbb{R}^{2,1}
      \times
      \,
    }
    \widehat X^{8}
    \ar@{}[ddr]|-{\mbox{\tiny (pb)}}
    \ar[r]_-{
      \underset{
        \mathclap{
        \;\;\;\;\;\;\;\;\;
        \color{greenii}
        \mbox{
          \tiny
          \begin{tabular}{c}
            dual C-field in
            \\
            J-twisted 7-Cohomotopy
          \end{tabular}
        }
        }
      }
      {
        c_6
      }
    }
    \ar[dd]|-{
      \mbox{
        \tiny
        \color{greenii}
        \begin{tabular}{c}
          3-sphere fibration
          \\
          over spacetime
        \end{tabular}
      }
    }
    &
    \overset{
      \mathrlap{
      \!\!\!\!\!\!\!\!\!\!
      \!\!\!\!\!\!\!\!\!\!
      \mbox{
        \tiny
        \color{darkblue}
        \begin{tabular}{c}
          classifying space for
          \\
          J-twisted 7-Cohomotopy
        \end{tabular}
      }
      }
    }{
      S^7 \!\sslash \mathrm{Sp}(2)
    }
    \ar@/^6pc/[rrrd]_<<<<<<<<<<<<{\ }="s2"
    |<<<<<<<<<<{
      \underset{
        \mbox{
          \tiny
          \begin{tabular}{c}
            $\phantom{a}$
            \\
            $\phantom{a}$
          \end{tabular}
        }
      }{
      \overset{
        \mathclap{
        \mbox{
          \tiny
          \color{greenii}
          \begin{tabular}{c}
            universal integral shifted C-field 4-flux
            \\
            universally restricted to M5-worldvolume
          \end{tabular}
        }
        }
      }{
          (h_{\mathbb{H}})^\ast \widetilde \Gamma_4
      }
      }
    }
    \ar[rr]|-{
      \overset{
        \mbox{
          \tiny
          $\phantom{a}$
        }
      }{
      \underset{
        \mathclap{
        \mbox{
          \tiny
          \color{greenii}
          coset space realization
        }
        }
      }{
        \simeq
      }
      }
    }
      ^<<<<<<<<<<<<<<<<<<<<<<{\ }="t2"
    \ar[dd]|-{
      \underset{
        \mathclap{
        \mbox{
          \tiny
          \color{greenii}
          \begin{tabular}{c}
            quaternionic
            \\
            Hopf fibration
          \end{tabular}
        }
        }
      }{
        h_{\mathbb{H}}
        \sslash
        \mathrm{Sp}(2)
      }
    }
    &&
    \overset{
      \mathllap{
      \mbox{
        \tiny
        \color{darkblue}
        \begin{tabular}{l}
          classifying space
          \\
          for gauge field on M5
        \end{tabular}
      }
      \;\;\;\;\;\;\;\;\;\;\;\;\;\;\;\;
      }
    }{
      B \mathrm{Sp}(1)_{L}
    }
    \ar[dr]|-{
      \mathllap{
        \mbox{
          \tiny
          \color{greenii}
          \begin{tabular}{c}
            $\phantom{a}$
            \\
            gauge
            \\
            instanton density
            \\
            (2nd Chern class)
          \end{tabular}
        }
        \!\!\!\!\!\!\!\!
      }
      c_2
    }
    \\
    \overset{
      \mathllap{
      \mbox{
        \tiny
        \color{darkblue}
        \begin{tabular}{c}
          (extended)
          \\
          M5-worldvolume
        \end{tabular}
      }
      \;\;\;\;\;\;
      }
    }{
      \widehat \Sigma
    }
    \ar@/^2.42pc/@{-->}[rrr]^>>>>>>>>>{
      \mbox{
        \tiny
        \color{purple}
        \begin{tabular}{c}
          induced
          \\
          non-abelian higher gauge field
          \\
          on M5-worldvolume
        \end{tabular}
      }
    }
    \ar[ur]|>>>>>>>>>>>{
      \mbox{
        \tiny
        \color{greenii}
        \begin{tabular}{c}
          B-field in
          \\
          twisted 3-Cohomotopy
        \end{tabular}
      }
      \;\;\;\;\;\;\;\;\;\;
    }
    ^<<<<<<<{b_2\!\!}
    \ar[dr]|>>>>>>>>>>>{
      {
      \mbox{
        \tiny
        \color{greenii}
        \begin{tabular}{c}
          embedding
          \\
          field
        \end{tabular}
      }
      \;\;\;
      }
    }
    _<<<<<<<{
      \phi \!\!\!
    }
    &
    &
    &
    \underset{
      \mathclap{
      \mbox{
        \tiny
        \color{darkblue}
        \begin{tabular}{c}
          classifying space for
          \\
          twisted String structure
        \end{tabular}
      }
      \;\;\;\;\;\;
      }
    }{
      B \mathrm{String}^{c_2}\!(4)
    }
    \ar@{}[rr]|>>>>>>>>>>>>>{\mbox{\tiny (pb)}}
    \ar[ur]_>>>>>>>>>>>{\ }="s"
    \ar[dr]^>>>>>>>>>>>{\ }="t"
    &
    &
    \overset{
      \mathclap{
      \mbox{
        \tiny
        \color{darkblue}
        \begin{tabular}{c}
          classifying space
          \\
          for
          \\
          circle 2-gerbes
        \end{tabular}
      }
      }
    }{
      B^3 U(1)
    }
    \\
    &
    \mathllap{
      \underset{
        \mathrlap{
          \mbox{
            \tiny
            \color{darkblue}
            \begin{tabular}{c}
              spacetime for
              \\
              M-theory on 8-mfds
            \end{tabular}
          }
        }
      }{
        \mathbb{R}^{2,1}
        \times
        \,
      }
    }
    X^8
    \ar[dr]_-{
      \underset{
        \mathclap{
        \mbox{
          \tiny
          \color{greenii}
          \begin{tabular}{c}
            tangent
            structure
          \end{tabular}
        }
        }
      }{
        T X^8
      }
    }
    \ar@/_3.925pc/[rrrr]|>>>>>>>>>>>>>>>>>>>>>>>{
      \overset{
        \mbox{
          \tiny
          $\phantom{a}$
        }
      }{
      \underset{
        \mathclap{
        \mbox{
          \tiny
          \color{greenii}
          shifted integral C-field 4-flux
        }
        }
      }{
        \; \widetilde G_4 \;\simeq_{{}_{\mathbb{R}}}\; G_4 + \frac{1}{4}p_1
      }
      }
    }
    \ar[r]^-{
      \overset{
        \mathclap{
        \;\;\;\;\;\;\;\;
        \mbox{
          \tiny
          \color{greenii}
          \begin{tabular}{c}
            C-field in
            \\
            J-twisted 4-Cohomotopy
          \end{tabular}
        }
        }
      }{
        c_3
      }
    }
    &
    \underset{
      \mathrlap{
      \!\!\!\!
      \mbox{
        \tiny
        \color{darkblue}
        \begin{tabular}{c}
          classifying space
          \\
          J-twisted 4-Cohomotopy
        \end{tabular}
      }
      }
    }{
      S^4 \!\sslash \mathrm{Spin}(5)
    }
    \ar@/_1.81pc/[rrr]|>>>>>>>>>>>>>>>>>{
      \overset{
        \mbox{
          \tiny
          \begin{tabular}{c}
            $\phantom{a}$
          \end{tabular}
        }
      }{
      \underset{
        \mathclap{
        \mbox{
          \tiny
          \color{greenii}
          \begin{tabular}{c}
            universal integral 4-flux
          \end{tabular}
        }
        \;\;\;\;\;
        }
      }{
        \widetilde \Gamma_4
      }
      }
    }
    \ar[d]
      |>>>>>>>{
        \phantom{ {AA} }
      }
    \ar[rr]|-{
      \overset{
        \mbox{
          \tiny
          $\phantom{a}$
        }
      }{
      \underset{
        \mathclap{
        \mbox{
          \tiny
          \color{greenii}
          \begin{tabular}{c}
            coset space realization
          \end{tabular}
        }
        }
      }{
        \simeq
      }
      }
    }
    &
    &
    \underset{
      \mathclap{
      \mbox{
        \tiny
        \color{darkblue}
        \begin{tabular}{c}
        \end{tabular}
      }
      }
    }{
      B \mathrm{Spin}(4)
    }
    \ar[ru]|-{
      \mathllap{
        \mbox{
          \tiny
          \color{greenii}
          \begin{tabular}{c}
            gravitational
            \\
            instanton density
            \\
            (\hspace{-.5pt}1\hspace{-.4pt}st \hspace{-.8pt}Pontrj. \hspace{-1pt}class)
          \end{tabular}
        }
        \!\!\!\!
      }
        \frac{1}{2}p_1
    }
    \ar[dll]
      |<<<<<<<<<<<<<<<<<<{
        \phantom{ {AA} \atop {{AA} \atop {AA}} }
      }
      |>>>>>>>>>>>>{
        \phantom{ {AA} \atop {AA} }
      }
    \ar[r]^-{
      \overset{
        \mathrlap{
          \;\;\;\;\;\;\;\;
          \overset{
            \!\!\!\!\!\!\!\!\!\!\!\!
            \mbox{
              \tiny
              \color{greenii}
              fractional
            }
          }{
          \mbox{
            \tiny
            \color{greenii}
            Euler + Pontrjagin class
          }
          }
        }
      }{
      \scalebox{.6}{
        $\tfrac{1}{2}\rchi_4 + \tfrac{1}{4}p_1$
      }
      }
    }
    &
    B^3 U(1)
    \\
    &
    &
    \underset{
      \mathclap{
      \mbox{
        \tiny
        \color{darkblue}
        \begin{tabular}{c}
          classifying space for
          \\
          Spin connection field
        \end{tabular}
      }
      }
    }{
      B \mathrm{Spin}(8)
    }
    &
    \ar@{=>}|-{
      \scalebox{.8}{
        $
        \underset{
          \mathclap{
          \mbox{
            \tiny
            \color{purple}
            \begin{tabular}{c}
              universal integral 3-flux
              \\
            \end{tabular}
          }
          }
        }{
          \mathclap{\phantom{\vert^{\vert^\vert}}}
          H^{{}^{\mathrm{univ}}}_3
        }
        $
      }
    } "s"; "t"
  }
  }
  };

 \draw (7.4,-3.05) node
 {
      \begin{tabular}{c}
        \begin{rotate}{-90}
          $\!\!\!\mathclap{\simeq}$
        \end{rotate}
        \\
        $
        \underset{
          \mbox{
            \tiny
            \color{darkblue}
            \begin{tabular}{c}
              classifying space for
              \\
              integral 4-cohomology
            \end{tabular}
          }
        }{
          \;\; K(\mathbb{Z},4)
        }
        $
      \end{tabular}
 };

 \draw (-2,5.1) node
 {
   \xymatrix@R=28pt@C=37pt{
     \overset{
       \mathclap{
       \mbox{
         \tiny
         \color{darkblue}
         \begin{tabular}{c}
           classifying space for
           \\
           J-twisted 7-cohomotopy
           \\
           with vanishing Euler 8-class
         \end{tabular}
       }
       }
     }{
       S^7 \!\sslash \widehat{\mathrm{Sp}(2)}
     }
     \ar[d]
     \ar[r]
     &
     \\
     &
   }
 };

 \draw (1.5,3.5) node
 {
   \xymatrix@R=25pt{
     \overset{
       \mathclap{
       \mbox{
         \tiny
         \color{darkblue}
         \begin{tabular}{c}
           classifying space for
           \\
           twisted String structure
           \\
           with vanishing Euler 8-class
         \end{tabular}
       }
       }
     }{
       B \mathrm{String}^{c_2}_{\rchi_8}\!(4)
     }
     \ar[dddd]
      |<<<<<<<<{
        \phantom{
          {A \atop } \atop { A \atop {A \atop A} }
        }
      }
      |>>>>>>>>>>>>>>>>>>>>>>{
        \phantom{{}_{\vert_{\vert_{\vert_{\vert_{\vert_{\vert}}}}}}}
      }
     &
     \\
     \\
     \\
     \\
     &
   }
 };

 \draw (8.3,4.9) node
 {
   \xymatrix@R=40pt@C=37pt{
     \overset{
       \mathclap{
       \mbox{
         \tiny
         \color{darkblue}
         \begin{tabular}{c}
           classifying space
           \\
           for
           \\
           circle 6-gerbes
         \end{tabular}
       }
       }
     }{
       B^7 U(1)
     }
     \ar@{}[d]
     &
     \\
     &
   }
 };

 \draw (-4.47,4.15) node
 {
   \xymatrix@C=46pt@R=54pt{
     &
     \\
     \ar@{..>}[ur]
   }
 };

 \draw[draw=white, fill=white] (-4.47,4.15) circle (.32);

 \draw (-4.47,4.15) node
 {
   \xymatrix@C=46pt@R=54pt{
     &
     \\
     \ar@{}[ur]
       |-{
       \mathclap{
       \;\;\;\;\;\;\;
       \mbox{
         \tiny
         \color{greenii}
         \begin{tabular}{c}
           Euler 8-class vanishes
           \\
           (M2-brane loci removed)
         \end{tabular}
       }
       }
     }
   }
 };

 \draw (4.45,4.8) node
 {
   \xymatrix@R=40pt@C=64pt{
     \ar@{}[d]
     \ar[rr]^-{
       \mbox{
         \tiny
         \color{purple}
         \begin{tabular}{c}
           stringy
           Hopf WZ-term of M5-brane
         \end{tabular}
       }
     }
     &&
     \\
     &
   }
 };

 \end{tikzpicture}
 }
 \hspace{1.1cm}

 \vspace{-.6cm}

\noindent

Here we show how the {\color{purple}purple} maps come about, and
what this means for the M5-brane. First some background:

\newpage

\noindent
{\bf The fundamental problem in string theory},
which remains open\footnote{
See, e.g.: \cite[6]{Duff96}\cite[p. 2]{HoweLambertWest97}\cite[p. 6]{Duff98}\cite[p. 2]{NicolaiHelling98}\cite[p. 330]{Duff99}\cite[12]{Moore14}\cite[p. 2]{ChesterPerlmutter18}\cite{Witten19}\cite{Duff19}.},
is to find its non-perturbative completion to
a conjectural {\it M-theory} \cite{Duff99}\cite{BeckerBeckerSchwarz06}.
This subsumes the major open problem
(e.g., \cite[p. 77]{Moore12}\cite[p. 49]{Lambert12}\cite[p. 1]{Hu13}\cite[3.6]{Lambert19})
of identifying the
{\it higher gauge theory}
(``non-abelian gerbe theory'' \cite[p. 6, 15]{Witten02}),
supposedly carried by
$N$ coincident M5-branes,
either seen as the 6d SCFT right on the brane
or seen in its vicinity (``AdS/CFT'', see \cite{Duff-singleton}) in the
non-perturbative {\it small $N$} limit
(large $1/N$ limit).
This is the problem we are concerned with here.
Specifically,
for $N \to 3$ this should, in turn, subsume
(``AdS/QCD'', e.g. \cite[\S 4]{RhoZahed16})
the Millennium Problem (see \cite{RobertsSchmidt20}\cite{CRoberts21})
of formulating non-perturbative YM theory,
hence confined QCD.

\vspace{1mm}
\noindent
{\bf The beauty of the M-theory conjecture} is that it reduces a
zoo of physical fields to just a pair:\footnote{Here we disregard the super-partner field -- the gravitino --
for focus of the exposition. In our context and under Hypothesis
H the full super-exceptional multiplet of M-theory fields is discussed in
\cite{FSS19d}\cite{FSS20a}.}
gravity and the C-field (review in \cite{MiemicSchnakenburg06}).
This suggests that formulating M-theory is
essentially tantamount to identifying the true mathematical nature
of the C-field coupled to gravity.
But this is tractable by an Ansatz like
\hyperlink{HypothesisH}{Hypothesis H},
because
at least the mathematical {\it type} of the C-field
is known with some certainty:

\vspace{1mm}
\noindent
{\bf The C-field is a higher gauge field}
with flux densities higher-degree differential forms \cite{SSS08},
and with gauge
transformations accompanied by higher-order gauge-of-gauge transformations
($n$-group gauge symmetry \cite{BaezSchreiber07}
\cite{FSS10}\cite{FSS13}\cite{dcct}).
Such higher gauge fields are described,
non-perturbatively,
by
{\it homotopy theory},
where they are embodied in
{\it non-abelian generalized cohomology}
theories (see \cite[\S 2]{FSS20b}). Coupling these to
other fields, notably to gravity, means to consider
{\it twisted} and {\it differential}
(see \eqref{DifferentialRefinement})
non-abelian generalized cohomology:

\vspace{1mm}
\label{CorrespondenceBetweenHigherGaugeFieldsAndHomotopyTheory}
\noindent
{\bf The correspondence between higher gauge fields,
generalized cohomology and geometric homotopy theory} proceeds as follows
\cite{SSS12}\cite{FSS12a}\cite{dcct}\cite{FSS20b}\cite{SS20b}\cite{SS20c}\cite{SS21}
(see \cite{FSS19a} for a gentle survey):

\vspace{3mm}
\hspace{-6mm}
{\small
\begin{tabular}{|c|c|c||p{2.5cm}|}
  \hline
  \begin{tabular}{c}
    Non-perturbative
    \\
    {\bf Physics}
  \end{tabular}
  &
  \begin{tabular}{c}
    Geometric (stacky)
    \\
    {\bf Homotopy theory}
  \end{tabular}
  &
  \begin{tabular}{c}
    Twisted differential
    \\
    {\bf Cohomology}
  \end{tabular}
  &
 \quad References
 \\
 \hline
 \hline
 Field configurations
 &
 $
   \raisebox{10pt}{
   \xymatrix@C=20pt{
     \underset{
       \mathclap{
       \raisebox{-3pt}{
         \tiny
         \color{darkblue}
         \bf
         spacetime
       }
       }
     }{
       X
     }
     \;\;
     \ar[rr]
       ^-{
         \mbox{
           \tiny
           \color{greenii}
           \bf
           maps
         }
       }
     &&
     \quad
     \underset{
       \mathclap{
       \raisebox{-3pt}{
         \tiny
         \color{darkblue}
         \bf
         \begin{tabular}{c}
           classifying
           \\
           space ($\infty$-stack)
         \end{tabular}
       }
       }
     }{
       A
     }
   }
   }
 $
 &
 cocycles
  &
  \multirow{4}{*}{
    \begin{minipage}[center]{3cm}
      {$\;$}
      \\
      \cite[Fig. 2]{SSS12}
      \\
      \cite[\S 3]{NSS12}
      \\
      \cite[\S 2.1]{FSS20b}
      \\
      \cite[(22)]{SS20b}
      \\
      \cite[\S 2.1]{SS21}
    \end{minipage}
  }
  \\
  \cline{1-3}
  \begin{tabular}{c}
    Gauge
    \\
    transformations
  \end{tabular}
  &
  $
    \xymatrix{
      X
      \ar@/^1.3pc/[rr]
        ^{}
        _-{\ }="s"
      \ar@/_1.3pc/[rr]
        _{}
        ^-{\ }="t"
      &&
      A
      \ar@{=>}
        |-{
          \mathclap{\phantom{\vert}}
          \mbox{
            \tiny
            \color{orangeii}
            \bf
            homotopies
          }
          \mathclap{\phantom{\vert}}
        }
        "s"; "t"
    }
  $
  &
  coboundaries
  &
  \\
  \cline{1-3}
  \begin{tabular}{c}
    Gauge-of-gauge
    \\
    transformations
  \end{tabular}
  &
  $
    \xymatrix{
      X
      \ar@/^1.3pc/[rr]
        ^-{}
        _-{\ }="s"
      \ar@/_1.3pc/[rr]
        _-{}
        ^-{\ }="t"
      &
      \mathclap{
      \scalebox{.8}{
        \raisebox{-2pt}{
        \begin{tikzpicture}
          \draw (0,-.04)
            node
            {
              \scalebox{1.5}{$\Rightarrow$}
            };
          \draw[line width=.56] (-.254,0) to (+.21,0);
        \end{tikzpicture}
        }
      }
      }
      &
      A
      \ar@/^1.1pc/@{=>}
        "s"; "t"
      \ar@/_1.1pc/@{=>}
        "s"; "t"
    }
  $
  &
  cocorners
  &
  \\
 \cline{1-3}
 \begin{tabular}{c}
   Gauge-equivalence classes
   \\
   of field configuration
 \end{tabular}
 &
 $
   A(X)
   \;\coloneqq\;
   \left\{
   \!\!\!\!
   \raisebox{2pt}{
   \xymatrix@C=9pt{
     X
     \ar[rr]
       ^-{
         \mbox{
           \tiny
           \color{greenii}
           \bf
           maps
         }
       }
     &&
     A
   }
   }
   \!\!\!
   \right\}
     _{
       \!\!\!\!
       \big/
         {
           \!\!\!\!\!
           \mbox{
             \tiny
             \color{orangeii}
             \bf
             homotopies
           }
         }
     }
 $
 &
 cohomology
  &
  \\
  \hline
  \hline
  \begin{tabular}{c}
    Field configurations
    \\
    coupled to gravity
  \end{tabular}
  &
  $
    \raisebox{20pt}{
    \xymatrix@R=14pt{
      X
      \ar[rr]
        |>>>>>>>>>>>>>>>>{
          \;
          \mbox{
            \tiny
            \color{greenii}
            \bf
            maps
          }
          \;
        }
        _-{\ }="s"
        \ar[dr]
          _<<<<<<<{
            \underset{
              \mathllap{
              \mbox{
                \tiny
                \color{greenii}
                \bf
                \begin{tabular}{c}
                  classifying map
                  \\
                  of frame bundle
                \end{tabular}
              }
              }
            }{
              T X
            }
          }
        ^>>>>{\ }="t"
      &
      \ar@{}[d]
        |<<{
          \mbox{
            \tiny
            \&
            $\mathclap{\phantom{\vert_{\vert}}}$
          }
        }
      &
      A \!\sslash\! \mathrm{Spin}(n)
      \ar[dl]
      \ar@{}[d]
        |<<{
          \raisebox{0pt}{
            \tiny
            \color{darkblue}
            \bf
            \begin{tabular}{c}
              homotopy
              \\
              quotient
            \end{tabular}
          }
        }
      \\
      &
      B \mathrm{Spin}(n)
      &
      \ar@{=>}
        ^-{
          \;\;\;
          \mbox{
            \tiny
            \color{orangeii}
            \bf
            homotopies
          }
        }
        "s"; "t"
    }
    }
  $
  &
  \begin{tabular}{c}
    tangentially
    \\
    twisted cocycle
  \end{tabular}
  &
  \raisebox{0pt}{
  \begin{minipage}[left]{3cm}
    \cite[\S 4]{NSS12}
    \\
    \cite[\S 2]{FSS19b}
    \\
    \cite[2.94]{SS20b}
    \\
    \cite[\S 2.2]{FSS20b}
    $\mathclap{\phantom{\vert_{\vert}}}$
  \end{minipage}
  }
  \\
 \hline
 \hline
 \begin{tabular}{c}
   Instanton/soliton
   \\
   numbers
 \end{tabular}
 &
 $
   \raisebox{10pt}{
   \xymatrix@C=17pt{
     \underset{
       \mathclap{
       \raisebox{-3pt}{
         \tiny
         \color{darkblue}
         \bf
         \begin{tabular}{c}
           classifying
           \\
           space ($\infty$-stack)
         \end{tabular}
       }
       }
     }{
       A
     }
     \ar[rr]
       ^-{
         \mbox{
           \tiny
           \color{greenii}
           \bf
           maps
         }
       }
       \quad
     &&
     \quad
     \underset{
       \mathclap{
       \raisebox{-3pt}{
         \tiny
         \color{darkblue}
         \bf
         \begin{tabular}{c}
           Eilenberg-MacLane
           \\
           space ($\infty$-stack)
         \end{tabular}
       }
       }
     }{
       B^n \mathrm{U}(1)
     }
   }
   }
 $
 &
 \begin{tabular}{c}
   characteristic
   \\
   classes
 \end{tabular}
 &
 \multirow{2}{*}{
   \begin{minipage}[left]{3cm}
     {$\;$}
     \\
     \cite{FSS10}
     \\
     \cite[\S 3.1]{FSS12a}
     \\
     \cite{FSS12c}
     \\
     \cite{FSS13}
   \end{minipage}
 }
 \\
 \cline{1-3}
 \begin{tabular}{c}
   Ordinary gauge instantons /
   \\
   gravitational instantons
 \end{tabular}
 &
 $
   \raisebox{17pt}{
   \xymatrix@C=12pt@R=-5pt{
     &&
     \underset{
       \mathclap{
       \raisebox{-3pt}{
       }
       }
     }{
       B \mathrm{U}
     }
     \ar[rr]
       _-{
         c_2
       }
     &&
     \underset{
       \mathclap{
       \raisebox{-3pt}{
         \tiny
         \color{darkblue}
         \bf
       }
       }
     }{
       B^3 \mathrm{U}(1)
     }
   \\
   X
   \ar[urr]
     ^-{
       \mbox{
         \tiny
         \color{greenii}
         \bf
         \begin{tabular}{c}
           gauge field
         \end{tabular}
       }
       \;
     }
   \ar[drr]
     _-{
       \mbox{
         \tiny
         \color{greenii}
         \bf
         gravity
       }
     }
   \\
     &&
     \underset{
       \mathclap{
       \raisebox{-3pt}{
         \tiny
         \color{darkblue}
         \bf
       }
       }
     }{
       B \mathrm{O}
     }
     \ar[rr]
       ^-{
         \scalebox{.5}{$\tfrac{1}{2}$}p_1
       }
     &&
     \underset{
       \mathclap{
       \raisebox{-3pt}{
         \tiny
         \color{darkblue}
         \bf
       }
       }
     }{
       B^3 \mathrm{U}(1)
     }
   }
   }
 $\vspace{-.32cm}
 &
 \begin{tabular}{c}
   2nd Chern class /
   \\
   1st Pontrjagin class
 \end{tabular}
 &
 \\
 \hline
 \hline
 \begin{tabular}{c}
   Green-Schwarz
   \\
   mechanism
 \end{tabular}
 &
 $
   \raisebox{20pt}{
   \xymatrix{
     B \mathrm{String}^{c_2}
     \ar[d]
       ^>>>{\ }="t"
     \ar[rr]
       _>>>{\ }="s"
     &&
     B \mathrm{U}
     \ar[d]
       ^-{ c_2 }
     \\
     B \mathrm{O}
     \ar[rr]
       _-{
         \scalebox{.5}{$\tfrac{1}{2}$}p_1
       }
     &&
     B^3 \mathrm{U}(1)
     \ar@{=>}
       _-{
         \mbox{
           \tiny \rm (pb)
         }
       }
       ^-{
         \mbox{
           \tiny
           \color{orangeii}
           \bf
           \begin{tabular}{c}
             homotopy
             \\
             pullback
           \end{tabular}
         }
       }
       "s"; "t"
   }
   }
 $
 &
 \begin{tabular}{c}
   secondary
   \\
   invariants
 \end{tabular}
 &
 \multirow{1}{*}{
 \begin{minipage}[left]{3cm}
   \cite{SSS12}
   \\
   \cite{FSS12a}
 \end{minipage}
 }
 \\
 \hline
\end{tabular}
}

\vspace{.2cm}

\noindent
These are the building blocks of Diagram
\eqref{TheFactorization}.
We now explain how this pertains to the M5-brane
gauge structure.


\medskip
\noindent {\bf The open problem of the gauge structure on the M5-brane}
(\cite[p. 6, 15]{Witten02}\cite[3.6]{Lambert19}) has two aspects to it:

\vspace{.1cm}

\hspace{-8.5mm}
\hypertarget{QuestionNonAbelianGaugeField}{}
\hypertarget{QuestionHigherGaugeField}{}
{\small
\begin{tabular}{|c||c|}
\hline
\begin{minipage}[left]{8.2cm}
\vspace{1mm}
{\bf Q. 1 -- {\color{darkblue}Non-abelian gauge structure} on M5-branes.}
\\
\emph{How does a non-abelian gauge group arise on coincident M5-branes?}
\\
(Given that the traditional
argument for non-abelian gauge fields on D-branes,
via perturbative open string scattering, does not apply.)
\end{minipage}
&
\begin{minipage}[left]{8.2cm}
\noindent {\bf Q. 2 -- {\color{darkblue}Higher gauge structure} on M5-branes.}
\\
\emph{How does this get promoted to a \emph{higher gauge group}
for a \emph{non-abelian gerbe gauge theory}?}
\\
(Given that the gauge potential on the M5-brane is, locally,
not a 1-form as in Yang-Mills theory, but a 2-form.)
\\
\phantom{A}
\end{minipage}
\\
\hline
\end{tabular}
}
\vspace{.1cm}

\vspace{.1cm}

\noindent
But the putative higher non-abelian
gauge field strength $H_3$ on the M5-brane is,
in the abelian case,
an incarnation on the brane of the
bulk $C_3$-field (see \cite[(4)]{BLNPST97}) --
for whose description we now invoke
\hyperlink{HypothesisH}{\it Hypothesis H}:

\vspace{.2cm}

\hspace{-.9cm}
\begin{tabular}{ll}
\begin{minipage}[left]{10.7cm}
\noindent {\bf Hypothesis H in the diagrammatic language}
from p. \pageref{CorrespondenceBetweenHigherGaugeFieldsAndHomotopyTheory}
says that the C-field in M-Theory on 8-manifolds is classified by
a map lifting the tangent bundle classifier
to the homotopy quotient of the (homotopy type of) the 4-sphere
by its canonical rotation action.
\end{minipage}
&
\hspace{4mm}
\raisebox{8pt}{
\scalebox{.9}{
\xymatrix@R=7pt@C=15pt{
  \mathbb{R}^{2,1}
  \times X^8
  \ar[rr]
    |-{
      \;c_3\;
    }
    ^-{
        \raisebox{0pt}{
          \tiny
          \color{greenii}
          \bf
          \begin{tabular}{c}
            M-theory C-field in
            \\
            J-twisted Cohomotopy
          \end{tabular}
        }
    }
    _>>>>>>>>{\phantom{-}}="s"
  \ar[dr]
    _-{T X^8}
    ^>>>{\ }="t"
  &&
  S^4 \!\sslash\! \mathrm{Spin}(5)
  \ar[dl]
  \\
  &
  B \mathrm{Spin}(8)
  \ar@{=>} "s"; "t"
}
}
}
\end{tabular}

\vspace{.14cm}

\noindent
This hypothesis is motivated from hidden structures found
in the ``brane scan'' \cite{AETW87}\cite{Duff88}
or rather ``brane bouquet'' \cite{FSS13b}\cite[2.1]{HSS18},
which were revealed by a re-analysis of the $\kappa$-symmetric super $p$-brane
sigma-models through the lens of super homotopy theory
\cite{FSS15}\cite{FSS16a} (reviewed in \cite[\S 7]{FSS19a}).
The hypothesis has since been checked to rigorously imply
a fair number of phenomena expected to be characteristic of M-theory
\cite{FSS19b}\cite{FSS19c}\cite{SS19a}\cite{SS19b}\cite{SS19c}\cite{SS20a}\cite{SS20c}\cite{SS21}
(exposition in \cite{Sc20}):

\vspace{.1cm}

\noindent {\bf The single heterotic M5-brane...}
In particular, we proved in \cite{FSS19d} that the
free Perry-Schwarz
action functional \cite{PerrySchwarz97}\cite{Schwarz97}\cite{APPS97} for the
single M5-brane in heterotic M-theory
(Ho{\v r}ava-Witten theory
\cite{HoravaWitten95}\cite{Witten96} \cite{HoravaWitten96}\cite{LLO97}\cite{LOW99}\cite{DOPW99}\cite{DOPW00}\cite{Ovrut02}\cite{Ovrut18})
\emph{emerges}
from \hyperlink{HypothesisH}{\it Hypothesis H}
(following an analogous but much simpler derivation
of the action functional for the single M2-brane in \cite[6.2]{HSS18}).

\vspace{.1cm}

\noindent {\bf ...is already non-abelian.}
But the \emph{heterotic} M5-brane is special in that
even a single such brane is expected to carry a non-abelian
(higher) gauge field, namely for gauge group the
quaternionic unitary group $\mathrm{Sp}(1) \simeq \mathrm{SU}(2)$.
This has been argued
by identifying heterotic
NS5-branes with ``small instantons'' in the
$\mathfrak{so}(32)$-heterotic gauge field \cite{Witten95}\cite{AspinwallGross96}\cite{AspinwallMorrison97},
and it has dually \cite{Sen99} been argued as an effect of gauge enhancement
for coincident 5-branes on orientifolds \cite{Mukhi97}:
D5-branes in type I string theory \cite{GimonPolchinski96},
M5-branes in heterotic M-theory  \cite{GanorHanany96}.

\vspace{1mm}
This second perspective says that
a single heterotic M5-brane may be understood as consisting of
two coincident M5-branes that are bound as mirror images of the
orientifold group action, thus revealing the
non-abelian $\mathrm{Sp}(1)$ gauge field on the brane as an instance of
the general phenomenon expected on coincident M5-branes.
Therefore it makes sense to investigate questions 
\hyperlink{QuestionNonAbelianGaugeField}{\bf Q. 1}
and
\hyperlink{QuestionHigherGaugeField}{\bf Q. 2}
already in the case of single heterotic M5-branes:

\vspace{2mm}

\hspace{-.7cm}
\begin{tikzpicture}[decoration=snake]

\draw (-.3,+.03) node {

\begin{tabular}{|c||c|c|c|}
    \hline
    \raisebox{-12pt}{
      \begin{tabular}{c}
      \bf Coincident M5s $\mathrlap{\to}$
      \\
      and their
      \\
      \bf worldvolume fields:
      \end{tabular}
    }
    &
    $N \!=\! \mathbf{1}$ $\mathrm{M5}$
    &
    \raisebox{3pt}{
    \xymatrix@R=7pt{
      N_{{}_{\mathrm{HET}}} \!=\! \mathbf{1} \; \mathrm{M5}_{{}_{\mathrm{HET}}}
      \ar@{=}[d]
      \\
      \;\mathrm{N} =\! \mathbf{2}/\mathbb{Z}_2 \; \mathrm{M5}
    }
    }
    &
    \color{gray}
    \raisebox{-23pt}{
      $N \gt \mathbf{2} \; \mathrm{M5}$
    }
    \\
    \hline
    \hline
    \begin{tabular}{c}
      Gauge field
    \end{tabular}
    &
    \begin{tabular}{c}
      none
    \end{tabular}
    &
    \begin{tabular}{c}
      $\mathrm{Sp}(1)$-bundle
    \end{tabular}
    &
    \color{gray} unclear
    \\
    \hline
    \begin{tabular}{c}
      Higher gauge field
    \end{tabular}
    &
    \begin{tabular}{c}
      $U(1)$-gerbe
      \\
      =
      $B U(1)$ 2-bundle
    \end{tabular}
    &
    \begin{tabular}{c}
      nonabelian gerbe:
      \\
      $\mathrm{String}^{c_2}\!(4)$ 2-bundle
    \end{tabular}
    &
    \color{gray} unclear
    \\
    \hline
\end{tabular}
};

 \draw[draw=green, fill=green, opacity=.25] (2.6,-.25) ellipse (1.2 and .3);
 \draw[draw=cyan, fill=cyan, opacity=.25] (2.6,-1.2) ellipse (1.7 and .4);

 \draw (8.6,.4) node
  {
    \scalebox{.85}{
    \it
    \begin{tabular}{c}
      argued in
      \\
      \cite{Witten96}\cite{AspinwallGross96}
      \\
      \cite{GimonPolchinski96}\cite{GanorHanany96}
    \end{tabular}
    }
  };

  \draw[green, opacity=.25, line width=2.4pt] (8.6-1.1,.4) to (2.6+1.1,-.25+.1);

 \draw (8.6,-.8-.3) node
  {
    \scalebox{.85}{
    \it
    \begin{tabular}{c}
      derived
      in Theorem \ref{GaugeFieldEmerges}
      \\
      from \hyperlink{HypothesisH}{\it Hypothesis H}
    \end{tabular}
    }
  };

  \draw[green, opacity=.25, line width=2.4pt] (8.6-1.5,-.8+.1-.3) to (2.6+1.1,-.25-.1);
  \draw[cyan, opacity=.25, line width=2.4pt] (8.6-1.5,-.8-.1-.3) to (2.6+1.7,-1.2);

\end{tikzpicture}

\vspace{1mm}

\noindent {\bf Results.}
We prove here that
\hyperlink{HypothesisH}{\it Hypothesis H}
implies,
for the topological sector of single heterotic M5-branes:

\vspace{-.4cm}

\begin{enumerate}[{\bf (1)}]
\setlength\itemsep{-4pt}
\item the emergence of a non-abelian $\mathrm{Sp}(1)$-gauge field
structure on the brane worldvolume;
\item the emergence of a lift to a higher non-abelian gauge field
(non-abelian gerbe field),
  specifically to a $\mathrm{String}^{c_2}$-structure on the M5-brane worldvolume
  (as in \cite[2.1]{Sati11}, generalizing the notion of \cite{CHZ});
\item  the worldvolume 3-flux $H_3$ being the corresponding non-abelian
Chern-Simons 2-gerbe \cite{FSS13}\cite{NSS12} (cf. \cite{CJMSW}\cite{Waldorf})
 reflecting a Green-Schwarz-type mechanism
on the worldvolume \cite{SSS12}\cite{FSS12a} spring
\end{enumerate}

\vspace{-.25cm}

\noindent All these results follow from inspection of the single encapsulating
homotopy-commutative diagram \eqref{TheFactorization}
in Theorem \ref{GaugeFieldEmerges},
using just basic homotopy theory
(see pointers in \cite{NSS12}\cite[\S A]{FSS20b}),
the results of \cite{FSS19b}\cite{FSS19c} and,
for interpreting the factorization as $\mathrm{String}^{c_2}$-structure,
the concepts developed in \cite{FSS12a},
as we now explain.

\newpage


\noindent First we explain the existence of the outer part of
diagram \eqref{TheFactorization}.
All statements, proofs and references that we appeal to here are
given in \cite{FSS19b}.

\medskip

\hspace{-.9cm}


\vspace{3mm}
\noindent   Here we restrict attention to the subgroup
  $\mathrm{Sp}(2) \;\simeq\; \mathrm{Sp}(2) \boldsymbol{\cdot} \mathbb{Z}_2
  \;\subset\; \mathrm{Sp}(2) \boldsymbol{\cdot} \mathrm{Sp}(1)$
  only for sake of exposition.

\vspace{4mm}

\hspace{-1cm}
%
          }
          }
        }{
          {H_3 =}
          \atop
          {\mathbf{CS}(\omega) - \mathbf{CS}(A)}
        }
        $
      }
    } "s"; "t"
    \ar@{=>}|-{
      \mbox{
        \tiny
        \color{purple}
        background charge
      }
    }
       "s2"; "t2"
  }
  }
   };

\end{tikzpicture}
}
}
\end{tabular}

\vspace{.4cm}

\noindent
Under this equivalence,
{\bf the integral cohomology ring}
$H^4(B \mathrm{Spin}(4); \mathbb{Z})
\simeq \mathbb{Z}[\tfrac{1}{2}p_1,
{\color{greenii}\tfrac{1}{2}\rchi_4 + \tfrac{1}{4}p_1}]$
is identified as containing the
\emph{shifted Euler class}
$\color{greenii} \widetilde \Gamma_4 := \tfrac{1}{2}\rchi_4 + \tfrac{1}{4}p_1$;
the plain Euler class $\tfrac{1}{2}\rchi_4$ is the fiberwise volume element on $S^4$.
\\
\noindent {\bf Hypothesis H identifies} the C-field flux
with the pullback of this class along
the given cohomotopy cocycle $c_3$.

\vspace{.3cm}

\noindent This way {\bf  the integral cohomology structure of $B \mathrm{Spin}(4)$
implies the shifted flux quantization of the C-field}:
$$
  {\color{greenii}
  \tfrac{1}{2}\rchi_4
  +
  \tfrac{1}{4}p_1
  }
  \;\in H^4(B \mathrm{Spin}(4); \mathbb{Z})
  \;\;\;\;\;\;\;
  \Leftrightarrow
  \;\;\;\;\;\;\;
  {\color{greenii}
  G_4 + \tfrac{1}{4}p_1
  }
  \;\in
  H^4( X^8; \mathbb{Z} )
  \,.
$$

\newpage

\hspace{-.9cm}
\begin{tabular}{ll}
\begin{minipage}[left]{9cm}
  \noindent
  {\bf The other homotopy equivalence} of classifying spaces
  $S^7 \sslash \mathrm{Sp}(2) \simeq B \mathrm{Sp}(1)$
  is similarly induced by the coset space realization of the
  7-sphere as $S^7 \simeq \mathrm{Sp}(2)/\mathrm{Sp}(1)$.
  This generalizes to the full equivariance group as the homotopy quotient
  $S^7 \!\sslash\! \big(\mathrm{Sp}(2)\boldsymbol{\cdot}\mathrm{Sp}(1)\big)
   \;\simeq\; B \big(\mathrm{Sp}(1)\boldsymbol{\cdot}\mathrm{Sp}(1)\big)$.

\end{minipage}
&
\hspace{.4cm}
\raisebox{-21pt}{
\scalebox{.8}
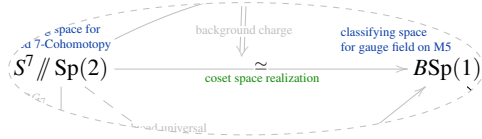
{
\begin{tikzpicture}

 \draw[dashed, lightgray] (-.35,3) ellipse (3.95 and 1.1);
 \clip (-.35,3) ellipse (3.95 and 1.1);

 \draw (0,0) node
 {
  \raisebox{68pt}{
  \xymatrix@C=35pt{
    \overset{
      \!\!\!\!\!\!\!\!
      \!\!\!\!\!\!\!\!
      \!\!\!\!
      \mathclap{
      \mbox{
        \tiny
        \color{darkblue}
        \begin{tabular}{c}
          classifying space for
          \\
          M5 sigma-model fields
        \end{tabular}
      }
    }
    }{
    \mathllap{
      \mathbb{R}^{2,1}
      \times
      \,
    }
    }
    \widehat X^{8}
    \ar@{}[ddr]|-{\mbox{\tiny (pb)}}
    \ar[r]_-{
      \mathclap{
      \mbox{
        \tiny
        \color{lightgray}
        \begin{tabular}{c}
          Hopf WZ-term
          \\
          $
          \;\;\;\;\;\;\;\;\;\;\;\;
          H_3
            \wedge
          (\widetilde G_4  - \tfrac{1}{2}p_1)
          +
          2 G_7
          $
        \end{tabular}
      }
      }
    }
    \ar[dd]|-{
      \mathclap{
      \mbox{
        \tiny
        \color{greenii}
        \begin{tabular}{c}
          embedding
          \\
          fields
        \end{tabular}
      }
      }
    }
    &
    \overset{
      \mathrlap{
      \!\!\!\!\!\!\!\!\!\!
      \!\!\!\!\!\!\!\!\!\!
      \mbox{
        \tiny
        \color{darkblue}
        \begin{tabular}{c}
          classifying space for
          \\
          J-twisted 7-Cohomotopy
        \end{tabular}
      }
      }
    }{
      S^7 \!\sslash \mathrm{Sp}(2)
    }
    \ar@[lightgray]@{-->}[dr]|-{
      \mbox{
        \tiny
        \color{lightgray}
        \begin{tabular}{c}
          induced universal
          \\
          non-abelian higher gauge field
          \\
          on M5-worldvolume
        \end{tabular}
      }
    }
    \ar@[lightgray]@/^6pc/[rrrd]_<<<<<<<<<{\ }="s2"
    |<<<<<<<<<<{
      \underset{
        \mbox{
          \tiny
          \begin{tabular}{c}
            $\phantom{a}$
            \\
            $\phantom{a}$
          \end{tabular}
        }
      }{
      \overset{
        \mathclap{
        \mbox{
          \tiny
          \color{greenii}
          \begin{tabular}{c}
            universal integral shifted C-field 4-flux
            \\
            universally restricted to M5-worldvolume
          \end{tabular}
        }
        }
      }{
          (h_{\mathbb{H}})^\ast \widetilde \Gamma_4
      }
      }
    }
    \ar@[lightgray][rr]|-{
      \overset{
        \mbox{
          \tiny
          $\phantom{a}$
        }
      }{
      \underset{
        \mathclap{
        \mbox{
          \tiny
          \color{greenii}
          coset space realization
        }
        }
      }{
        \simeq
      }
      }
    }
      ^<<<<<<<<<<<<<<<<<<<<<{\ }="t2"
    \ar@[lightgray][dd]|-{
      \underset{
        \mathclap{
        \mbox{
          \tiny
          \color{greenii}
          \begin{tabular}{c}
            Borel-equivariant
            \\
            quaternionic
            \\
            Hopf fibration
          \end{tabular}
        }
        }
      }{
        h_{\mathbb{H}}
        \sslash
        \mathrm{Sp}(2)
      }
    }
    &&
    \overset{
      \mathllap{
      \mbox{
        \tiny
        \color{darkblue}
        \begin{tabular}{l}
          classifying space
          \\
          for gauge field on M5
        \end{tabular}
      }
      \;\;\;\;\;\;\;\;\;\;\;\;\;\;\;\;
      }
    }{
      B \mathrm{Sp}(1)
    }
    \ar[dr]|-{
      \mathllap{
        \mbox{
          \tiny
          \color{greenii}
          \begin{tabular}{c}
            $\phantom{a}$
            \\
            gauge
            \\
            instanton density
            \\
            (2nd Chern class)
          \end{tabular}
        }
        \!\!\!\!\!\!\!\!
      }
      c_2
    }
    \\
    &
    &
    \underset{
      \mathclap{
      \mbox{
        \tiny
        \color{darkblue}
        \begin{tabular}{c}
          classifying space for
          \\
          twisted String structure
        \end{tabular}
      }
      \;\;\;\;\;\;
      }
    }{
      B \mathrm{String}^{c_2}\!(4)
    }
    \ar@{}[rr]|>>>>>>>>>>>>>{\mbox{\tiny (pb)}}
    \ar@[lightgray][ur]_>>>>>>>>>>{\ }="s"
    \ar[dr]^-{\ }="t"
    &
    &
    \overset{
      \mathclap{
      \mbox{
        \tiny
        \color{darkblue}
        \begin{tabular}{c}
          classifying space
          \\
          for
          \\
          circle 2-gerbes
        \end{tabular}
      }
      }
    }{
      B^3 U(1)
    }
    \\
    \mathllap{
      \underset{
        \mathrlap{
          \mbox{
            \tiny
            \color{darkblue}
            \begin{tabular}{c}
              spacetime for
              \\
              M-theory on 8-mfds
            \end{tabular}
          }
        }
      }{
        \mathbb{R}^{2,1}
        \times
        \,
      }
    }
    X^8
    \ar[dr]_-{
      \underset{
        \mathclap{
        \mbox{
          \tiny
          \color{greenii}
          \begin{tabular}{c}
            tangent
            structure
          \end{tabular}
        }
        }
      }{
        T X^8
      }
    }
    \ar@/_3.95pc/[rrrr]|>>>>>>>>>>>>>>>>>>>>>>>{
      \overset{
        \mbox{
          \tiny
          $\phantom{a}$
        }
      }{
      \underset{
        \mathclap{
        \mbox{
          \tiny
          \color{greenii}
          shifted integral C-field 4-flux
        }
        }
      }{
        \widetilde G_4 \;\simeq_{{}_{\mathbb{R}}}\; G_4 + \tfrac{1}{4}p_1
      }
      }
    }
    \ar[r]^-{
      \overset{
        \mathclap{
        \;\;\;\;\;\;\;\;
        \mbox{
          \tiny
          \color{greenii}
          \begin{tabular}{c}
            C-field in
            \\
            J-twisted Cohomotopy
          \end{tabular}
        }
        }
      }{
        c
      }
    }
    &
    \underset{
      \mathrlap{
      \!\!\!\!
      \mbox{
        \tiny
        \color{darkblue}
        \begin{tabular}{c}
          classifying space
          \\
          J-twisted 4-Cohomotopy
        \end{tabular}
      }
      }
    }{
      S^4 \!\sslash \mathrm{Spin}(5)
    }
    \ar@/_1.85pc/[rrr]|>>>>>>>>>>>>>>>>>{
      \overset{
        \mbox{
          \tiny
          \begin{tabular}{c}
            $\phantom{a}$
          \end{tabular}
        }
      }{
      \underset{
        \mathclap{
        \mbox{
          \tiny
          \color{greenii}
          \begin{tabular}{c}
            universal integral 4-flux
          \end{tabular}
        }
        \;\;\;\;\;
        }
      }{
        \widetilde \Gamma_4
      }
      }
    }
    \ar[d]
      |>>>>>>>{
        \phantom{ {AA} }
      }
    \ar[rr]|-{
      \overset{
        \mbox{
          \tiny
          $\phantom{a}$
        }
      }{
      \underset{
        \mathclap{
        \mbox{
          \tiny
          \color{greenii}
          \begin{tabular}{c}
            coset space realization
          \end{tabular}
        }
        }
      }{
        \simeq
      }
      }
    }
    &
    &
    \underset{
      \mathclap{
      \mbox{
        \tiny
        \color{darkblue}
        \begin{tabular}{c}
        \end{tabular}
      }
      }
    }{
      B \mathrm{Spin}(4)
    }
    \ar[ru]|-{
      \mathllap{
        \mbox{
          \tiny
          \color{greenii}
          \begin{tabular}{c}
            gravitational
            \\
            instanton density
            \\
            (\hspace{-.5pt}1\hspace{-.4pt}st \hspace{-.8pt}Pontrj. \hspace{-1pt}class)
          \end{tabular}
        }
        \!\!\!\!\!
      }
        \tfrac{1}{2}p_1
    }
    \ar[dll]
      |<<<<<<<<<<<<<<<<{
        \phantom{ {AA} \atop {{AA} \atop {AA}} }
      }
      |>>>>>>>>>>{
        \phantom{ {AA} \atop {AA} }
      }
    \ar[r]^-{
      \overset{
        \mathrlap{
          \;\;\;\;\;\;\;\;
          \overset{
            \!\!\!\!\!\!\!\!\!\!\!\!
            \mbox{
              \tiny
              \color{greenii}
              fractional
            }
          }{
          \mbox{
            \tiny
            \color{greenii}
            Euler + Pontrjagin class
          }
          }
        }
      }{
      \scalebox{.6}{
        $\tfrac{1}{2}\rchi_4 + \tfrac{1}{4}p_1$
      }
      }
    }
    &
    B^3 U(1)
    \mathrlap{
      \;
      {
      \overset{
        \mathclap{
        \scalebox{.5}{
          hmtpy
        }
        }
      }{
        \simeq
      }
      }
      \underset{
        \mathclap{
        \mbox{
          \tiny
          \color{darkblue}
          \begin{tabular}{c}
            classfying space for
            \\
            integral 4-cohomology
          \end{tabular}
        }
        }
      }{
        \, K(\mathbb{Z},4)
      }
    }
    \\
    &
    \underset{
      \mathclap{
      \mbox{
        \tiny
        \color{darkblue}
        \begin{tabular}{c}
          classifying space for
          \\
          M2-brane background structure
        \end{tabular}
      }
      }
    }{
      B \mathrm{Spin}(8)
    }
    &
    \ar@{=>}|-{
      \scalebox{.8}{
        $
        \underset{
          \mathclap{
          \mbox{
            \tiny
            \color{purple}
            \begin{tabular}{c}
              Green-Schwarz mechanism
            \end{tabular}
          }
          }
        }{
          {H_3 =}
          \atop
          {\mathbf{CS}(\omega) - \mathbf{CS}(A)}
        }
        $
      }
    } "s"; "t"
    \ar@[lightgray]@{=>}|-{
      \mbox{
        \tiny
        \color{lightgray}
        background charge
      }
    }
       "s2"; "t2"
  }
  }
 };

\end{tikzpicture}
}
}
\end{tabular}


\vspace{4mm}
\hspace{-1cm}
\begin{tabular}{ll}

\hspace{0cm}
\begin{minipage}[left]{6.6cm}
  {\bf The homotopy-commutativity of the}
  $\mathrlap{\mbox{\bf outer square}}$
  \\
  is seen as follows:
  \\
  With the exceptional isomorphisms
  \begin{equation}
    \label{Spin4IsSp1TimesSp1}
   \hspace{-.5cm}
    \begin{aligned}
      \mathrm{Spin}(4)
      &
      \;\simeq\;
      \mathrm{Spin}(3) \times \mathrm{Spin}(3)
      \\
      &
      \;\simeq\; \;\;\mathrm{Sp}(1)\;\, \times \;\;\mathrm{Sp}(1)\;\,
    \end{aligned}
  \end{equation}
  and the induced cohomology identifications
  \begin{equation}
    \begin{aligned}
    H^\bullet\big( B \mathrm{Sp}(1); \mathbb{Z} \big)
    & \;\simeq\;
    \mathbb{Z}\big[ c_2 \big]
    \\
    H^\bullet\big( B \mathrm{Spin}(4); \mathbb{Z} \big)
    & \;\simeq\;
    \mathbb{Z}\big[
      c^{\scalebox{.55}{$L$}}_2,
      c^{\scalebox{.55}{$R$}}_2
    \big]
    \end{aligned}
  \end{equation}
  the universal Pontrjagin class $\tfrac{1}{2}p_1$ on
  $B \mathrm{Spin}(4)$
  is identified with the sum of the
\end{minipage}

&
\scalebox{.8}{
\raisebox{-3.3cm}{
\begin{tikzpicture}

 \draw[dashed, gray] (-.1,.47) ellipse (6.4 and 3.5);
 \clip (-.1,.5) ellipse (6.4 and 3.5);

 \draw (0,0) node
  {
  \xymatrix@C=35pt{
    \overset{
      \!\!\!\!\!\!\!\!
      \!\!\!\!\!\!\!\!
      \!\!\!\!
      \mathclap{
      \mbox{
        \tiny
        \color{darkblue}
        \begin{tabular}{c}
          classifying space for
          \\
          M5 sigma-model fields
        \end{tabular}
      }
    }
    }{
    \mathllap{
      \mathbb{R}^{2,1}
      \times
      \,
    }
    }
    \widehat X^{8}
    \ar@{}[ddr]|-{\mbox{\tiny \color{lightgray} (pb)}}
    \ar@[lightgray][r]_-{
      \mathclap{
      \mbox{
        \tiny
        \begin{tabular}{c}
          \color{lightgray}
          Hopf WZ-term
          \\
          $
          \;\;\;\;\;\;\;\;\;\;\;\;
          H_3
            \wedge
          (\widetilde G_4  - \tfrac{1}{2}p_1)
          +
          2 G_7
          $
        \end{tabular}
      }
      }
    }
    \ar[dd]|-{
      \mathclap{
      \mbox{
        \tiny
        \color{lightgray}
        \begin{tabular}{c}
          embedding
          \\
          fields
        \end{tabular}
      }
      }
    }
    &
    \overset{
      \mathrlap{
      \!\!\!\!\!\!\!\!\!\!
      \!\!\!\!\!\!\!\!\!\!
      \mbox{
        \tiny
        \color{lightgray}
        \begin{tabular}{c}
          classifying space for
          \\
          J-twisted 7-Cohomotopy
        \end{tabular}
      }
      }
    }{
      S^7 \!\sslash \mathrm{Sp}(2)
    }
    \ar@[lightgray]@{-->}[dr]|-{
      \mbox{
        \tiny
        \color{lightgray}
        \begin{tabular}{c}
          induced universal
          \\
          non-abelian higher gauge field
          \\
          on M5-worldvolume
        \end{tabular}
      }
    }
    \ar@[lightgray]@/^6pc/[rrrd]_<<<<<<<<<{\ }="s2"
    |<<<<<<<<<<{
      \underset{
        \mbox{
          \tiny
          \begin{tabular}{c}
            $\phantom{a}$
            \\
            $\phantom{a}$
          \end{tabular}
        }
      }{
      \overset{
        \mathclap{
        \mbox{
          \tiny
          \color{greenii}
          \begin{tabular}{c}
            universal integral shifted C-field 4-flux
            \\
            universally restricted to M5-worldvolume
          \end{tabular}
        }
        }
      }{
          (h_{\mathbb{H}})^\ast \widetilde \Gamma_4
      }
      }
    }
    \ar[rr]|-{
      \overset{
        \mbox{
          \tiny
          $\phantom{a}$
        }
      }{
      \underset{
        \mathclap{
        \mbox{
          \tiny
          \color{greenii}
          coset space realization
        }
        }
      }{
        \simeq
      }
      }
    }
      ^<<<<<<<<<<<<<<<<<<<<<{\ }="t2"
    \ar[dd]|-{
      \underset{
        \mathclap{
        \mbox{
          \tiny
          \color{greenii}
          \begin{tabular}{c}
            Borel-equivariant
            \\
            quaternionic
            \\
            Hopf fibration
          \end{tabular}
        }
        }
      }{
        h_{\mathbb{H}}
        \sslash
        \mathrm{Sp}(2)
      }
    }
    &&
    \overset{
      \mathllap{
      \mbox{
        \tiny
        \color{lightgray}
        \begin{tabular}{l}
          classifying space
          \\
          for gauge field on M5
        \end{tabular}
      }
      \;\;\;\;\;\;\;\;\;\;\;\;\;\;\;\;
      }
    }{
      B \mathrm{Sp}(1)
    }
    \ar[dr]|-{
      \mathllap{
        \mbox{
          \tiny
          \color{lightgray}
          \begin{tabular}{c}
            $\phantom{a}$
            \\
            gauge
            \\
            instanton density
            \\
            ({\color{greenii}2nd Chern class})
          \end{tabular}
        }
        \!\!\!\!\!\!\!\!
      }
      c_2
    }
    \\
    &
    &
    \underset{
      \mathclap{
      \mbox{
        \tiny
        \color{lightgray}
        \begin{tabular}{c}
          classifying space for
          \\
          twisted String structure
        \end{tabular}
      }
      \;\;\;\;\;\;
      }
    }{
      {\color{lightgray} B \mathrm{String}^{c_2}\!(4)}
    }
    \ar@{}[rr]|>>>>>>>>>>>>>{\mbox{\tiny \color{lightgray}(pb)}}
    \ar@[lightgray][ur]_>>>>>>>>>>{\ }="s"
    \ar@[lightgray][dr]^-{\ }="t"
    &
    &
    \overset{
      \mathclap{
      \mbox{
        \tiny
        \color{lightgray}
        \begin{tabular}{c}
          classifying space
          \\
          for
          \\
          circle 2-gerbes
        \end{tabular}
      }
      }
    }{
      B^3 U(1)
    }
    \\
    \mathllap{
      \underset{
        \mathrlap{
          \mbox{
            \tiny
            \color{darkblue}
            \begin{tabular}{c}
              spacetime for
              \\
              M-theory on 8-mfds
            \end{tabular}
          }
        }
      }{
        \mathbb{R}^{2,1}
        \times
        \,
      }
    }
    X^8
    \ar[dr]_-{
      \underset{
        \mathclap{
        \mbox{
          \tiny
          \color{greenii}
          \begin{tabular}{c}
            tangent
            structure
          \end{tabular}
        }
        }
      }{
        T X^8
      }
    }
    \ar@/_3.95pc/[rrrr]|>>>>>>>>>>>>>>>>>>>>>>>{
      \overset{
        \mbox{
          \tiny
          $\phantom{a}$
        }
      }{
      \underset{
        \mathclap{
        \mbox{
          \tiny
          \color{greenii}
          shifted integral C-field 4-flux
        }
        }
      }{
        \widetilde G_4 \;\simeq_{{}_{\mathbb{R}}}\; G_4 + \tfrac{1}{4}p_1
      }
      }
    }
    \ar[r]^-{
      \overset{
        \mathclap{
        \;\;\;\;\;\;\;\;
        \mbox{
          \tiny
          \color{lightgray}
          \begin{tabular}{c}
            C-field in
            \\
            J-twisted Cohomotopy
          \end{tabular}
        }
        }
      }{
        c
      }
    }
    &
    \underset{
      \mathrlap{
      \!\!\!\!
      \mbox{
        \tiny
        \color{lightgray}
        \begin{tabular}{c}
          classifying space
          \\
          J-twisted 4-Cohomotopy
        \end{tabular}
      }
      }
    }{
      S^4 \!\sslash \mathrm{Spin}(5)
    }
    \ar@[lightgray]@/_1.85pc/[rrr]|>>>>>>>>>>>>>>>>>{
      \overset{
        \mbox{
          \tiny
          \begin{tabular}{c}
            $\phantom{a}$
          \end{tabular}
        }
      }{
      \underset{
        \mathclap{
        \mbox{
          \tiny
          \color{greenii}
          \begin{tabular}{c}
            universal integral 4-flux
          \end{tabular}
        }
        \;\;\;\;\;
        }
      }{
        \widetilde \Gamma_4
      }
      }
    }
    \ar[d]
      |>>>>>>>{
        \phantom{ {AA} }
      }
    \ar[rr]|-{
      \overset{
        \mbox{
          \tiny
          $\phantom{a}$
        }
      }{
      \underset{
        \mathclap{
        \mbox{
          \tiny
          \color{greenii}
          \begin{tabular}{c}
            coset space realization
          \end{tabular}
        }
        }
      }{
        \simeq
      }
      }
    }
    &
    &
    \underset{
      \mathclap{
      \mbox{
        \tiny
        \color{darkblue}
        \begin{tabular}{c}
        \end{tabular}
      }
      }
    }{
      B \mathrm{Spin}(4)
    }
    \ar[ru]|-{
      \mathllap{
        \mbox{
          \tiny
          \color{lightgray}
          \begin{tabular}{c}
            gravitational
            \\
            instanton density
            \\
            ({\color{greenii}\hspace{-.5pt}1\hspace{-.4pt}st \hspace{-.8pt}Pontrj. \hspace{-1pt}class})
          \end{tabular}
        }
        \!\!\!\!\!
      }
        \tfrac{1}{2}p_1
    }
    \ar@[lightgray][dll]
      |<<<<<<<<<<<<<<<<{
        \phantom{ {AA} \atop {{AA} \atop {AA}} }
      }
      |>>>>>>>>>>{
        \phantom{ {AA} \atop {AA} }
      }
    \ar@[lightgray][r]^-{
      \overset{
        \mathrlap{
          \;\;\;\;\;\;\;\;
          \overset{
            \!\!\!\!\!\!\!\!\!\!\!\!
            \mbox{
              \tiny
              \color{lightgray}
              fractional
            }
          }{
          \mbox{
            \tiny
            \color{lightgray}
            Euler + Pontrjagin class
          }
          }
        }
      }{
      \scalebox{.6}{
        \color{lightgray}
        $\tfrac{1}{2}\rchi_4 + \tfrac{1}{4}p_1$
      }
      }
    }
    &
    B^3 U(1)
    \mathrlap{
      \;
      {
      \overset{
        \mathclap{
        \scalebox{.5}{
          hmtpy
        }
        }
      }{
        \simeq
      }
      }
      \underset{
        \mathclap{
        \mbox{
          \tiny
          \color{darkblue}
          \begin{tabular}{c}
            classfying space for
            \\
            integral 4-cohomology
          \end{tabular}
        }
        }
      }{
        \, K(\mathbb{Z},4)
      }
    }
    \\
    &
    \underset{
      \mathclap{
      \mbox{
        \tiny
        \color{darkblue}
        \begin{tabular}{c}
          classifying space for
          \\
          M2-brane background structure
        \end{tabular}
      }
      }
    }{
      B \mathrm{Spin}(8)
    }
    &
    \ar@[lightgray]@{=>}|-{
      \scalebox{.8}{
        $
        \underset{
          \mathclap{
          \mbox{
            \tiny
            \color{lightgray}
            \begin{tabular}{c}
              Green-Schwarz mechanism
            \end{tabular}
          }
          }
        }{
          {
          \color{lightgray}
          {H_3 =}
          \atop
          {\mathbf{CS}(\omega) - \mathbf{CS}(A)}
          }
        }
        $
      }
    } "s"; "t"
    \ar@[lightgray]@{=>}|-{
      \mbox{
        \tiny
        \color{lightgray}
        background charge
      }
    }
       "s2"; "t2"
  }
  };

\end{tikzpicture}
}
}
\end{tabular}

\vspace{.07cm}

\noindent
  second Chern classes of
the two $\mathrm{Sp}(1)$-factors (by \cite[Lemma 3.9]{FSS19b}),
  and
  the Borel-equivariant quaternionic Hopf fibration
  is identified with the inclusion of the left
  $\mathrm{Sp}(1)$-factors (by \cite[Prop. 2.22]{FSS19b}),
  as shown here:
  \begin{equation}
    \label{TheComponents}
    \raisebox{78pt}{
    \xymatrix@R=5pt@C=6pt{
      S^7 \!\sslash\! \mathrm{Sp}(2)
      \ar[dddd]|-{
        \mathllap{
          \mbox{
            \tiny
            \color{greenii}
            \begin{tabular}{c}
              Borel-equivariant
              \\
              quaternionic
              \\
              Hopf fibration
            \end{tabular}
          }
          \;\;
        }
        h_{\mathbb{H}} \sslash \mathrm{Sp}(2)
      }
      &
      \ar@{<->}[rr]^{\simeq}_-{
        \mathclap{
        \mbox{
          \tiny
          \color{greenii}
          \begin{tabular}{c}
            coset space
            \\
            realization
          \end{tabular}
        }
        }
      }
      &&
      &
      B \mathrm{Sp}(1)
      \ar[dddd]
      &
      \ar@{<->}[rr]^-{=}
      &&
      &
      B \mathrm{Sp}(1)_L
      \ar@{}[r]|-{\mbox{$\times$}}
      \ar@{=}[dddd]|-{
        \mathclap{
        \mbox{
          \tiny
          \color{greenii}
          \begin{tabular}{c}
            inclusion of
            \\
            left factor
          \end{tabular}
        }
        }
      }
      &
      \;\;\;\;\;\ast\;\;\;\;\;
      \ar[dddd]
      \\
      \\
      &&&& &&&& &&&&
      \mathrlap{
      \!\!\!\!\!\!
      \scalebox{.8}{\cite[2.22]{FSS19b}}
      }
      \\
      \\
      S^4 \!\sslash\! \mathrm{Spin}(5)
      &
      \ar@{<->}[rr]_-{\simeq}^-{
        \mathclap{
        \mbox{
          \tiny
          \color{greenii}
          \begin{tabular}{c}
            coset space
            \\
            realization
          \end{tabular}
        }
        }
      }
      &&
      &
      B \mathrm{Spin}(4)
      &
      \ar@{<->}[rr]_-{\simeq}^-{
        \mathclap{
        \mbox{
          \tiny
          \color{greenii}
          \begin{tabular}{c}
            exceptional
            \\
            isomorphism
          \end{tabular}
        }
        }
      }
      &&
      &
      B \mathrm{Sp}(1)_L
      \ar@{}[r]|-{\mbox{$\times$}}
      &
      B \mathrm{Sp}(1)_R
      \\
      \\
      &
      &&
      &
      H^4(B \mathrm{Spin}(4); \mathbb{Z})
      &
      \ar@{<->}[rr]^{\simeq}
      &&
      &
      H^4(B \mathrm{Sp}(1); \mathbb{Z})
      \ar@{}[r]|-{\mbox{$\oplus$}}
      &
      H^4(B \mathrm{Sp}(1); \mathbb{Z})
      \\
      &
      &&
      &
      \mathllap{
        \mbox{
          \tiny
          \color{darkblue}
          \begin{tabular}{c}
            first Pontrjagin class/
            \\
            C-field background charge
          \end{tabular}
        }
      }
      \tfrac{1}{2}p_1
      &&\mathclap{=}&&
      c^{\scalebox{.55}{$L$}}_2
      \ar@{}[r]|-{\mbox{$+$}}
      &
      c^{\scalebox{.55}{$R$}}_2
      \mathrlap{
        \;\;\;
        \mbox{
          \tiny
          \color{darkblue}
          total Chern class
        }
      }
      \\
      &
      &&
      &
      \mathllap{
        \mbox{
          \tiny
          \color{darkblue}
          \begin{tabular}{c}
            shifted fractional Euler class/
            \\
            shifted integral C-field flux
          \end{tabular}
        }
        \;
        \widetilde \Gamma_4
        \;=\;
      }
      \tfrac{1}{2}\rchi_4 + \tfrac{1}{4}p_1
      &&\mathclap{=}&&
      c^{\scalebox{.55}{$L$}}_2
      \mathrlap{
        \mbox{
          \tiny
          \color{darkblue}
          \begin{tabular}{c}
            left Chern class
          \end{tabular}
        }
      }
      &
      &&
      \mathrlap{
        \;\;\;
        \scalebox{.8}{\cite[3.9]{FSS19b}}
      }
      \\
      &
      &&
      &
      \mathllap{
        \mbox{
          \tiny
          \color{darkblue}
          \begin{tabular}{c}
            C-field flux relative
            \\
            to background charge
          \end{tabular}
        }
      }
      \widetilde \Gamma_4 - \tfrac{1}{2}p_1
      &&\mathclap{=}&&
      &
      \mathllap{-}
      c^{\scalebox{.55}{$R$}}_2
      \mathrlap{
        \mbox{
          \tiny
          \color{darkblue}
          \begin{tabular}{c}
            right Chern class
          \end{tabular}
        }
      }
    }
    }
  \end{equation}
This makes it manifest that the pullback of
the Pontrjagin class along the Borel-equivariant quaternionic Hopf fibration
kills the right Chern class summand and retains the left Chern class
summand, which is again the plain Chern class on the single
$\mathrm{Sp}(1)$-factor that is the domain of the
Borel-equivariant quaternionic Hopf fibration:
\begin{equation}
  \label{CommutativityAsEquation}
  (h_{\mathbb{H}}\sslash \mathrm{Sp}(2))^\ast\
  \big(
    \tfrac{1}{2}p_1
  \big)
  \;=\;
  c_2
  \,.
\end{equation}
Since $B^3 U(1) \simeq K(\mathbb{Z},4)$,
this equality of cohomology classes comes from
a homotopy between their representative maps to classifying spaces,
and this is the homotopy commutativity of the outer square above.

\medskip

\hspace{-.9cm}



\vspace{3mm}
We now explain the factorization through twisted String structure
in the inner part of the diagram \eqref{TheFactorization}.
The terminology and conceptualization of twisted String structures
and their relation to the Green-Schwarz mechanism \cite{GreenSchwarz}
follows \cite{Sati11}\cite{SSS12}\cite{FSS12a}\cite{FSS12b}.

\medskip

\hspace{-1cm}
%
 }
 }
\end{tabular}

\vspace{3mm}

\noindent This being a \emph{homotopy}-fiber product means that its defining
square diagram is filled by a
universal homotopy $H^{{}^{\mathrm{univ}}}_3\!\!\!\!$,
as shown above,
that is a coboundary between
cocycle representatives of these two characteristic classes
after their pullback to
the homotopy-fiber product space $B \mathrm{String}^{c_2}\!(4)$.
As such,
$H^{{}^{\mathrm{univ}}}_3\!\!\!\!$
is the homotopy-theoretic reflection of the
\emph{Green-Schwarz mechanism} \cite{SSS12}:
Upon enhancing all classifying spaces of topological structures
in \eqref{HomotopyFiberProduct}
to their corresponding \emph{moduli stacks} of differential structures
(according to \cite{FSS10}\cite{SSS12}\cite{FSS12a}\cite{FSS12b},
see \eqref{DifferentialRefinement} below)
this homotopy is given, locally, by a 3-form flux $H_3$ whose
de Rham differential is the heterotic Bianchi identity:

\vspace{-.6cm}

\begin{equation}
  \label{HeteroticBianchi}
  d
  \underset{
    \mathclap{
    \mbox{
      \tiny
      \color{darkblue}
      \begin{tabular}{c}
        3-flux form
      \end{tabular}
    }
    }
  }{
    \underbrace{
      H_3
    }
  }
    \;=\;
  \underset{
    \mathclap{
    \mbox{
      \tiny
      \begin{tabular}{c}
        $= \tfrac{1}{2}p_1(R)$
        \\
        \color{darkblue}
        Pontrjagin form
      \end{tabular}
    }
    }
  }{
    \underbrace{
      \mathrm{Tr}(R \wedge R)
    }
  }
    -
  \underset{
    \mathclap{
    \mbox{
      \tiny
      \begin{tabular}{c}
        $= c_2(F)$
        \\
        \color{darkblue}
        Chern form
      \end{tabular}
    }
    }
  }{
  \underbrace{
    \mathrm{Tr}(F \wedge F)
  }
  }\;.
\end{equation}
Here $R$ and $F$ denote, as usual, the local curvature forms of
connections on the given $\mathrm{Spin}(4)$- and $\mathrm{Sp}(1)$-bundle,
respectively, and we take the traces to be integrally normalized.

\hspace{-.9cm}
\begin{tabular}{ll}
\begin{minipage}[left]{11cm}
\noindent {\bf The dashed factorization} in the inner part of
\eqref{TheFactorization} through the homotopy fiber product
\eqref{HomotopyFiberProduct},
shown on the right equivalently with domain $S^7 \!\sslash \mathrm{Sp}(2)$,
is now the immediate consequence of
the commutativity of the outer square, as established above \eqref{CommutativityAsEquation},
due to the defining universal property of homotopy-fiber products.

\vspace{.1cm}

\noindent {\bf In conclusion}, this shows that $\mathrm{Sp}(1)$ gauge- and
$\mathrm{String}^{c_2}(4)$ higher gauge structure emerges
whenever a cocycle in J-twisted 4-Cohomotopy factors through
the classifying space $S^7 \!\sslash\! \mathrm{Sp}(2)$ of
J-twisted 7-Cohomotopy via the Borel-equivariant quaternionic Hopf fibration.

\end{minipage}
&
\raisebox{-65pt}{
\scalebox{.9}{
\begin{tikzpicture}

 \draw[dashed, gray, rotate=-37] (-2.55,.3) ellipse (3.5 and 1.4);
 \clip[rotate=-37] (-2.55,.3) ellipse (3.5 and 1.4);

\draw (0,0) node
{
  \raisebox{68pt}{
  \xymatrix@C=35pt{
    \overset{
      \!\!\!\!\!\!\!\!
      \!\!\!\!\!\!\!\!
      \!\!\!\!
      \mathclap{
      \mbox{
        \tiny
        \color{darkblue}
        \begin{tabular}{c}
          classifying space for
          \\
          M5 sigma-model fields
        \end{tabular}
      }
    }
    }{
    \mathllap{
      \mathbb{R}^{2,1}
      \times
      \,
    }
    }
    \widehat X^{8}
    \ar@{}[ddr]|-{\mbox{\tiny (pb)}}
    \ar@[lightgray][r]_-{
      \mathclap{
      \mbox{
        \tiny
        \color{lightgray}
        \begin{tabular}{c}
          Hopf WZ-term
          \\
          $
          \;\;\;\;\;\;\;\;\;\;\;\;
          H_3
            \wedge
          (\widetilde G_4  - \tfrac{1}{2}p_1)
          +
          2 G_7
          $
        \end{tabular}
      }
      }
    }
    \ar[dd]|-{
      \mathclap{
      \mbox{
        \tiny
        \color{greenii}
        \begin{tabular}{c}
          embedding
          \\
          fields
        \end{tabular}
      }
      }
    }
    &
    \overset{
      \mathrlap{
      \!\!\!\!\!\!\!\!\!\!
      \!\!\!\!\!\!\!\!\!\!
      \mbox{
        \tiny
        \color{darkblue}
        \begin{tabular}{c}
          classifying space for
          \\
          J-twisted 7-Cohomotopy
        \end{tabular}
      }
      }
    }{
      S^7 \!\sslash \mathrm{Sp}(2)
    }
    \ar@{-->}[dr]|-{
      \mbox{
        \tiny
        \color{purple}
        \begin{tabular}{c}
          induced universal
          \\
          non-abelian higher gauge field
          \\
          on M5-worldvolume
        \end{tabular}
      }
    }
    \ar@[lightgray]@/^6pc/[rrrd]_<<<<<<<<<{\ }="s2"
    |<<<<<<<<<<{
      \underset{
        \mbox{
          \tiny
          \begin{tabular}{c}
            $\phantom{a}$
            \\
            $\phantom{a}$
          \end{tabular}
        }
      }{
      \overset{
        \mathclap{
        \mbox{
          \tiny
          \color{greenii}
          \begin{tabular}{c}
            universal integral shifted C-field 4-flux
            \\
            universally restricted to M5-worldvolume
          \end{tabular}
        }
        }
      }{
          (h_{\mathbb{H}})^\ast \widetilde \Gamma_4
      }
      }
    }
    \ar@[lightgray][rr]|-{
      \overset{
        \mbox{
          \tiny
          $\phantom{a}$
        }
      }{
      \underset{
        \mathclap{
        \mbox{
          \tiny
          \color{lightgray}
          coset space realization
        }
        }
      }{
        \simeq
      }
      }
    }
      ^<<<<<<<<<<<<<<<<<<<<<{\ }="t2"
    \ar@[lightgray][dd]|-{
      \underset{
        \mathclap{
        \mbox{
          \tiny
          \color{greenii}
          \begin{tabular}{c}
            quaternionic
            \\
            Hopf fibration
          \end{tabular}
        }
        }
      }{
        {\color{lightgray}h_{\mathbb{H}}
        \sslash
        \mathrm{Sp}(2)
        }
      }
    }
    &&
    \overset{
      \mathllap{
      \mbox{
        \tiny
        \color{darkblue}
        \begin{tabular}{l}
          classifying space
          \\
          for gauge field on M5
        \end{tabular}
      }
      \;\;\;\;\;\;\;\;\;\;\;\;\;\;\;\;
      }
    }{
      B \mathrm{Sp}(1)
    }
    \ar[dr]|-{
      \mathllap{
        \mbox{
          \tiny
          \color{greenii}
          \begin{tabular}{c}
            $\phantom{a}$
            \\
            gauge
            \\
            instanton density
            \\
            (2nd Chern class)
          \end{tabular}
        }
        \!\!\!\!\!\!\!\!
      }
      c_2
    }
    \\
    &
    &
    \underset{
      \mathclap{
      \mbox{
        \tiny
        \color{darkblue}
        \begin{tabular}{c}
          classifying space for
          \\
          twisted String structure
        \end{tabular}
      }
      \;\;\;\;\;\;
      }
    }{
      B \mathrm{String}^{c_2}\!(4)
    }
    \ar@{}[rr]|>>>>>>>>>>>>>{\mbox{\tiny (pb)}}
    \ar@[lightgray][ur]_>>>>>>>>>>{\ }="s"
    \ar[dr]^-{\ }="t"
    &
    &
    \overset{
      \mathclap{
      \mbox{
        \tiny
        \color{darkblue}
        \begin{tabular}{c}
          classifying space
          \\
          for
          \\
          circle 2-gerbes
        \end{tabular}
      }
      }
    }{
      B^3 U(1)
    }
    \\
    \mathllap{
      \underset{
        \mathrlap{
          \mbox{
            \tiny
            \color{darkblue}
            \begin{tabular}{c}
              spacetime for
              \\
              M-theory on 8-mfds
            \end{tabular}
          }
        }
      }{
        \mathbb{R}^{2,1}
        \times
        \,
      }
    }
    X^8
    \ar[dr]_-{
      \underset{
        \mathclap{
        \mbox{
          \tiny
          \color{greenii}
          \begin{tabular}{c}
            tangent
            structure
          \end{tabular}
        }
        }
      }{
        T X^8
      }
    }
    \ar@/_3.95pc/[rrrr]|>>>>>>>>>>>>>>>>>>>>>>>{
      \overset{
        \mbox{
          \tiny
          $\phantom{a}$
        }
      }{
      \underset{
        \mathclap{
        \mbox{
          \tiny
          \color{greenii}
          shifted integral C-field 4-flux
        }
        }
      }{
        \widetilde G_4 \;\simeq_{{}_{\mathbb{R}}}\; G_4 + \tfrac{1}{4}p_1
      }
      }
    }
    \ar[r]^-{
      \overset{
        \mathclap{
        \;\;\;\;\;\;\;\;
        \mbox{
          \tiny
          \color{greenii}
          \begin{tabular}{c}
            C-field in
            \\
            J-twisted Cohomotopy
          \end{tabular}
        }
        }
      }{
        c
      }
    }
    &
    \underset{
      \mathrlap{
      \!\!\!\!
      \mbox{
        \tiny
        \color{darkblue}
        \begin{tabular}{c}
          classifying space
          \\
          J-twisted 4-Cohomotopy
        \end{tabular}
      }
      }
    }{
      S^4 \!\sslash \mathrm{Spin}(5)
    }
    \ar@/_1.85pc/[rrr]|>>>>>>>>>>>>>>>>>{
      \overset{
        \mbox{
          \tiny
          \begin{tabular}{c}
            $\phantom{a}$
          \end{tabular}
        }
      }{
      \underset{
        \mathclap{
        \mbox{
          \tiny
          \color{greenii}
          \begin{tabular}{c}
            universal integral 4-flux
          \end{tabular}
        }
        \;\;\;\;\;
        }
      }{
        \widetilde \Gamma_4
      }
      }
    }
    \ar[d]
      |>>>>>>>{
        \phantom{ {AA} }
      }
    \ar[rr]|-{
      \overset{
        \mbox{
          \tiny
          $\phantom{a}$
        }
      }{
      \underset{
        \mathclap{
        \mbox{
          \tiny
          \color{greenii}
          \begin{tabular}{c}
            coset space realization
          \end{tabular}
        }
        }
      }{
        \simeq
      }
      }
    }
    &
    &
    \underset{
      \mathclap{
      \mbox{
        \tiny
        \color{darkblue}
        \begin{tabular}{c}
        \end{tabular}
      }
      }
    }{
      B \mathrm{Spin}(4)
    }
    \ar[ru]|-{
      \mathllap{
        \mbox{
          \tiny
          \color{greenii}
          \begin{tabular}{c}
            gravitational
            \\
            instanton density
            \\
            (\hspace{-.5pt}1\hspace{-.4pt}st \hspace{-.8pt}Pontrj. \hspace{-1pt}class)
          \end{tabular}
        }
        \!\!\!\!\!
      }
        \tfrac{1}{2}p_1
    }
    \ar[dll]
      |<<<<<<<<<<<<<<<<{
        \phantom{ {AA} \atop {{AA} \atop {AA}} }
      }
      |>>>>>>>>>>{
        \phantom{ {AA} \atop {AA} }
      }
    \ar[r]^-{
      \overset{
        \mathrlap{
          \;\;\;\;\;\;\;\;
          \overset{
            \!\!\!\!\!\!\!\!\!\!\!\!
            \mbox{
              \tiny
              \color{greenii}
              fractional
            }
          }{
          \mbox{
            \tiny
            \color{greenii}
            Euler + Pontrjagin class
          }
          }
        }
      }{
      \scalebox{.6}{
        $\tfrac{1}{2}\rchi_4 + \tfrac{1}{4}p_1$
      }
      }
    }
    &
    B^3 U(1)
    \mathrlap{
      \;
      {
      \overset{
        \mathclap{
        \scalebox{.5}{
          hmtpy
        }
        }
      }{
        \simeq
      }
      }
      \underset{
        \mathclap{
        \mbox{
          \tiny
          \color{darkblue}
          \begin{tabular}{c}
            classfying space for
            \\
            integral 4-cohomology
          \end{tabular}
        }
        }
      }{
        \, K(\mathbb{Z},4)
      }
    }
    \\
    &
    \underset{
      \mathclap{
      \mbox{
        \tiny
        \color{darkblue}
        \begin{tabular}{c}
          classifying space for
          \\
          M2-brane background structure
        \end{tabular}
      }
      }
    }{
      B \mathrm{Spin}(8)
    }
    &
    \ar@{=>}|-{
      \scalebox{.8}{
        $
        \underset{
          \mathclap{
          \mbox{
            \tiny
            \color{purple}
            \begin{tabular}{c}
              Green-Schwarz mechanism
            \end{tabular}
          }
          }
        }{
          {H_3 =}
          \atop
          {\mathbf{CS}(\omega) - \mathbf{CS}(A)}
        }
        $
      }
    } "s"; "t"
    \ar@{=>}|-{
      \mbox{
        \tiny
        \color{purple}
        background charge
      }
    }
       "s2"; "t2"
  }
  }
};

\end{tikzpicture}
}
}

\end{tabular}

\newpage

\label{TheFieldContent}
Next we explain, with \cite{FSS15}\cite{FSS19c}\cite{FSS19d}, how this space
$S^7  \!\sslash\! \mathrm{Sp}(2)$ is,
under \hyperlink{HypothesisH}{\it Hypothesis H},
the classifying space
for the higher gauge field in the M5-brane sigma-model
with target space $\mathbb{R}^{2,1} \times X^8$.

\vspace{.5cm}

\hspace{-.9cm}
\begin{tabular}{ccc}
\fcolorbox{lightgray}{white}{
\hspace{-.52cm}
\scalebox{.8}{
\raisebox{-21pt}{
\xymatrix@C=13pt@R=22pt{
  \overset{
    \mathclap{
    \mbox{
      \tiny
      \color{darkblue}
      \begin{tabular}{c}
        exceptional
        \\
        super-spacetime
      \end{tabular}
    }
    }
  }{
    \mathbb{T}^{10,1\vert \mathbf{32}}_{\mathrm{exc},s}
  }
  \ar[ddddr]|<<<<<<<<<<{
    \mbox{
      \tiny
      \color{greenii}
      \begin{tabular}{c}
        rational
        \\
        super 517-torus fibration
        \\
        over super-spacetime
      \end{tabular}
    }
  }
  \ar[r]
  &
  \overset{
    \mathclap{
    \mbox{
      \tiny
      \color{darkblue}
      \begin{tabular}{c}
        extended
        \\
        super-spacetime
      \end{tabular}
    }
    }
  }{
    \widehat {\mathbb{T}^{10,1\vert \mathbf{32}}}
  }
  \ar@{}[ddddrrrr]|-{
    \mbox{
      \tiny
      (pb)
    }
  }
  \ar[dddd]|-{
    \mathclap{
    \mbox{
      \tiny
      \color{greenii}
      \begin{tabular}{c}
        rational
        \\
        3-sphere fibration
        \\
        over super-spacetime
      \end{tabular}
    }
    }
  }
  \ar[rrrr]^-{
    \overset{
      \mbox{
        \tiny
        \color{greenii}
        \begin{tabular}{c}
          super WZ term of M5
        \end{tabular}
      }
    }{
      \overbrace{
        \scalebox{.7}{
          $
          h_3 \wedge \mu_{M2}
          \;+\;
          \mu_{M5}
          $
        }
      }
    }
  }
  &&&&
  S^7_{\mathbb{R}}
  \ar[dddd]|-{
    \underset{
      \mbox{
        \tiny
        \color{greenii}
        \begin{tabular}{c}
          rationalized
          \\
          quaternionic
          \\
          Hopf fibration
        \end{tabular}
      }
    }{
    \big(
      h_{\mathbb{H}}
    \big)_{\mathbb{R}}
    }
  }
  \\
  \\
  \\
  \\
  &
  \overset{
    \mbox{
      \tiny
      \color{darkblue}
      \begin{tabular}{c}
        $D = 11$, $\mathcal{N}=1$
        \\
        super-spacetime
      \end{tabular}
    }
  }{
    \mathbb{T}^{10,1\vert \mathbf{32}}
  }
  \!\!\!\!\!\!\!\!\!
  \ar[rrrr]_-{
    \scalebox{.6}{
    $
    \underset{
      \;\;\;\;\;\;\;\;\;\;\;\;\;\;\;\;\;\;
      \scalebox{1.6}{
        \tiny
        \color{greenii}
        \begin{tabular}{c}
          completion to
          \\
          cocycle in (rational super) 4-Cohomotopy
        \end{tabular}
      }
    }{
    \overset{
      \scalebox{1.6}{
        \tiny
        \color{greenii}
        \begin{tabular}{c}
          super WZ term of M2
        \end{tabular}
      }
    }{
    \left(
      \begin{aligned}
      \mu_{M2}
        & :=
      \overbrace{
      \tfrac{i}{2}
      \big(
        \overline \psi
        \Gamma_{a_1 a_2}
        \psi
      \big)
      \,
      e^{a_1} \wedge e^{a_2}
      }
      \,,
      \\
      \mu_{M5}
        & :=
      \underbrace{
      \tfrac{1}{5}
      \big(
        \overline \psi
        \Gamma_{a_1 \cdots a_5}
        \psi
      \big)
      \,
      e^{a_1} \wedge \cdots \wedge e^{a_5}
      }
      \end{aligned}
    \right)
    }
    }
    $
    }
  }
  &&&&
  S^4_{\mathbb{R}}
}
}
}
\hspace{-.53cm}
}
&
\;\;
\multirow{2}{*}{
\hspace{-.55cm}
\begin{minipage}[left]{5.8cm}
  \noindent
  {\bf The 3-sphere fibration over spacetime}
  associated with the given cocycle $c_3$ in
  J-twisted Cohomotopy theory is the homotopy-pullback
  of the Borel-equivariant quaternionic Hopf fibration
  along $c_3$. This is the direct analog in fibered topological spaces
  of the construction in rational super-spaces
  (survey in \cite{FSS19a})
  which
  induces the super WZ term of the M5-brane sigma model \cite{FSS15},
  exhibiting 3-spherical T-duality of the M5 \cite{FSS18} \cite{SS18},
  and from that the full $\kappa$-symmetric Lagrangian
  of the M5-sigma model \cite{FSS19d}.
\end{minipage}
\hspace{-.4cm}
}
&
\!\!\!
\raisebox{-181pt}{
\scalebox{.8}{

}
}
\\
{\color{olive} \bf In rational super-spaces}
&&
{\color{olive} \bf In topological spaces}
\end{tabular}

\medskip
\medskip
\hspace{-.9cm}
%
 }
 }

 \end{tabular}

\medskip
\medskip
 \noindent {\bf This is the field content of the M5-brane sigma model}
 with worldvolume $\widehat{\Sigma}$ and target space
 $\mathbb{R}^{2,1}\!\!\times X^8$. \hspace{-2pt}Thus
 $\mathbb{R}^{2,1}\times \widehat{X}^8$ is identified with the
 classifying space for M5-brane sigma-model fields, for given
 background C-field.

\newpage

Now we are in position to state the main result of the present article:

\noindent {If the worldvolume field $\phi$ is indeed an embedding of Spin manifolds},
so that $\phi^\ast \tfrac{1}{2}p_1(T X^8) = \tfrac{1}{2}p_1(T \widehat \Sigma)$,
then {\bf the integral lift of the trivialization \eqref{BianchiOfH3OnSigma},
is a  \emph{twisted String structure} on the M5-brane} \cite[2.1]{Sati11}
namely a trivialization of $\tfrac{1}{2}p_1$ relative to
a background 4-class $\phi^\ast\widetilde G_4$ in integral cohomology.

\medskip
Here with \hyperlink{HypothesisH}{\it Hypothesis H}
we see further substructure in this phenomenon, in that
the pullback of the shifted integral C-field flux to the M5 worldvolume
is identified with the second Chern class of
an $\mathrm{Sp}(1)$-gauge field \eqref{TheComponents}
$\phi^\ast\widetilde G_4 \simeq c_2$, whence we
have specifically $c_2$-twisted String structure on the worldvolume.
In direct analogy with
$\mathrm{Spin}^c$-structure,
this is called \emph{$\mathrm{String}^{c_2}$-structure} \cite[2.1]{Sati11},
with classifying space $B \mathrm{String}^{c_2}$ \eqref{HomotopyFiberProduct}.

\vspace{.3cm}

\hspace{-.85cm}
\begin{tabular}{ll}
 \begin{minipage}[left]{7.45cm}

{\bf In conclusion}, the above discussion shows:
\vspace{-3mm}
\begin{enumerate}[{\bf (i)}]
\item  the homotopy-commutativity
of the total outer
part  of the diagram \eqref{TheFactorization};
\vspace{-3mm}
\item
the existence of the dashed factorization map
 \eqref{HomotopyFiberProduct}
induced by
the universal property of the homotopy fiber product $B \mathrm{String}^{c_2}$,
which identifies the field content \eqref{M5BraneFields}
of the M5-brane sigma model
with that of an embedding field together with a higher
$\mathrm{Sp}(1)$-gauge field with gauge 2-group $\mathrm{String}^{c_2}$.
\end{enumerate}
\vspace{-3mm}
 \end{minipage}
 &
 \raisebox{-32pt}{
 \scalebox{.8}
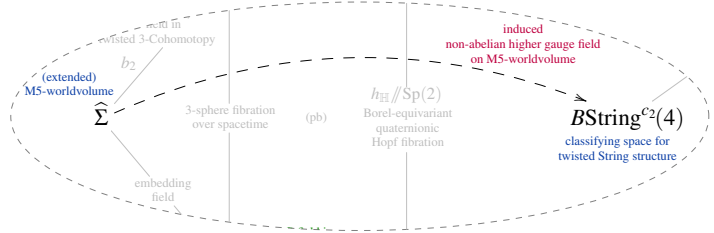
{
 \begin{tikzpicture}

  \draw[dashed, gray] (-3.5,0.5) ellipse (5.8 and 1.9);
  \clip (-3.5,0.5) ellipse (5.8 and 1.9);

  \draw (0,0) node
  {
  \raisebox{68pt}{
  \xymatrix@C=45pt{
    &
    \mathllap{
      \mathllap{
      \mbox{
        \tiny
        \color{darkblue}
        \begin{tabular}{c}
          3-sphere fiber
          \\
          (over any point)
        \end{tabular}
      }
      }
    }
    \;
    S^3
    \mathrlap{
      \;\,\simeq\;
      \mathrm{Sp}(1)_R
    }
    \ar[d]
    \\
    &
    \mathllap{
      \overset{
        \mbox{
          \tiny
          \color{darkblue}
          classifying space for
        }
      }
      {
        \mbox{
          \tiny
          \color{darkblue}
          M5 sigma-model fields
        }
      }
      \;
      \mathbb{R}^{2,1}
      \times
      \,
    }
    \widehat X^{8}
    \ar@{}[ddr]|-{\mbox{\color{lightgray}\tiny (pb)}}
    \ar[r]_-{
      \mathclap{
      \mbox{
        \tiny
        \begin{tabular}{c}
          \color{greenii}
          Hopf WZ-term
        \end{tabular}
      }
      }
    }
    \ar@[lightgray][dd]|-{
      \mbox{
        \tiny
        \color{lightgray}
        \begin{tabular}{c}
          3-sphere fibration
          \\
          over spacetime
        \end{tabular}
      }
    }
    &
    \overset{
      \mathrlap{
      \!\!\!\!\!\!\!\!\!\!
      \!\!\!\!\!\!\!\!\!\!
      \mbox{
        \tiny
        \color{darkblue}
        \begin{tabular}{c}
          classifying space for
          \\
          J-twisted 7-Cohomotopy
        \end{tabular}
      }
      }
    }{
      S^7 \!\sslash \mathrm{Sp}(2)
    }
    \ar@/^6pc/[rrrd]_<<<<<<<<<<<<{\ }="s2"
    |<<<<<<<<<<<<{
      \underset{
        \mbox{
          \tiny
          \begin{tabular}{c}
            $\phantom{a}$
            \\
            $\phantom{a}$
          \end{tabular}
        }
      }{
      \overset{
        \mathclap{
        \mbox{
          \tiny
          \color{greenii}
          \begin{tabular}{c}
            universal integral shifted C-field 4-flux
            \\
            universally restricted to M5-worldvolume
          \end{tabular}
        }
        }
      }{
          (h_{\mathbb{H}})^\ast \widetilde \Gamma_4
      }
      }
    }
    \ar[rr]|-{
      \overset{
        \mbox{
          \tiny
          $\phantom{a}$
        }
      }{
      \underset{
        \mathclap{
        \mbox{
          \tiny
          \color{greenii}
          coset space realization
        }
        }
      }{
        \simeq
      }
      }
    }
      ^<<<<<<<<<<<<<<<<<<<<<<{\ }="t2"
    \ar@[lightgray][dd]|-{
      \underset{
        \mathclap{
        \mbox{
          \tiny
          \color{lightgray}
          \begin{tabular}{c}
            Borel-equivariant
            \\
            quaternionic
            \\
            Hopf fibration
          \end{tabular}
        }
        }
      }{
        {\color{lightgray}h_{\mathbb{H}}
        \sslash
        \mathrm{Sp}(2)}
      }
    }
    &&
    \overset{
      \mathllap{
      \mbox{
        \tiny
        \color{darkblue}
        \begin{tabular}{l}
          classifying space
          \\
          for gauge field on M5
        \end{tabular}
      }
      \;\;\;\;\;\;\;\;\;\;\;\;\;\;\;\;
      }
    }{
      B \mathrm{Sp}(1)_{L}
    }
    \ar[dr]|-{
      \mathllap{
        \mbox{
          \tiny
          \color{greenii}
          \begin{tabular}{c}
            $\phantom{a}$
            \\
            gauge
            \\
            instanton density
            \\
            (2nd Chern class)
          \end{tabular}
        }
        \!\!\!\!\!\!\!\!
      }
      c_2
    }
    \\
    \overset{
      \mathllap{
      \mbox{
        \tiny
        \color{darkblue}
        \begin{tabular}{c}
          (extended)
          \\
          M5-worldvolume
        \end{tabular}
      }
      \;\;\;\;\;\;
      }
    }{
      \widehat \Sigma
    }
    \ar@/^2.42pc/@{-->}[rrr]^>>>>>>>>>{
      \mbox{
        \tiny
        \color{purple}
        \begin{tabular}{c}
          induced
          \\
          non-abelian higher gauge field
          \\
          on M5-worldvolume
        \end{tabular}
      }
    }
    \ar@[lightgray][ur]|>>>>>>>>>>>{
      \mbox{
        \tiny
        \color{lightgray}
        \begin{tabular}{c}
          B-field in
          \\
          twisted 3-Cohomotopy
        \end{tabular}
      }
      \;\;\;\;\;\;\;\;\;\;
    }
    ^<<<<<<<{\color{lightgray}b_2\!\!}
    \ar@[lightgray][dr]|-{
      {
      \mbox{
        \tiny
        \color{lightgray}
        \begin{tabular}{c}
          embedding
          \\
          field
        \end{tabular}
      }
      }
    }
    &
    &
    &
    \underset{
      \mathclap{
      \mbox{
        \tiny
        \color{darkblue}
        \begin{tabular}{c}
          classifying space for
          \\
          twisted String structure
        \end{tabular}
      }
      \;\;\;\;\;\;
      }
    }{
      B \mathrm{String}^{c_2}\!(4)
    }
    \ar@{}[rr]|>>>>>>>>>>>>>{\mbox{\tiny (pb)}}
    \ar@[lightgray][ur]_>>>>>>>>>>>{\ }="s"
    \ar[dr]^>>>>>>>>>>>{\ }="t"
    &
    &
    \overset{
      \mathclap{
      \mbox{
        \tiny
        \color{darkblue}
        \begin{tabular}{c}
          classifying space
          \\
          for
          \\
          circle 2-gerbes
        \end{tabular}
      }
      }
    }{
      B^3 U(1)
    }
    \\
    &
    \mathllap{
      \underset{
        \mathrlap{
          \mbox{
            \tiny
            \color{darkblue}
            \begin{tabular}{c}
              spacetime for
              \\
              M-theory on 8-mfds
            \end{tabular}
          }
        }
      }{
        \mathbb{R}^{2,1}
        \times
        \,
      }
    }
    X^8
    \ar[dr]_-{
      \underset{
        \mathclap{
        \mbox{
          \tiny
          \color{greenii}
          \begin{tabular}{c}
            tangent
            structure
          \end{tabular}
        }
        }
      }{
        T X^8
      }
    }
    \ar@/_3.925pc/[rrrr]|>>>>>>>>>>>>>>>>>>>>>>>{
      \overset{
        \mbox{
          \tiny
          $\phantom{a}$
        }
      }{
      \underset{
        \mathclap{
        \mbox{
          \tiny
          \color{greenii}
          shifted integral C-field 4-flux
        }
        }
      }{
        \; \widetilde G_4 \;\simeq_{{}_{\mathbb{R}}}\; G_4 + \frac{1}{4}p_1
      }
      }
    }
    \ar[r]^-{
      \overset{
        \mathclap{
        \;\;\;\;\;\;\;\;
        \mbox{
          \tiny
          \color{greenii}
          \begin{tabular}{c}
            C-field in
            \\
            J-twisted 4-Cohomotopy
          \end{tabular}
        }
        }
      }{
        c
      }
    }
    &
    \underset{
      \mathrlap{
      \!\!\!\!
      \mbox{
        \tiny
        \color{darkblue}
        \begin{tabular}{c}
          classifying space
          \\
          J-twisted 4-Cohomotopy
        \end{tabular}
      }
      }
    }{
      S^4 \!\sslash \mathrm{Spin}(5)
    }
    \ar@/_1.81pc/[rrr]|>>>>>>>>>>>>>>>>>{
      \overset{
        \mbox{
          \tiny
          \begin{tabular}{c}
            $\phantom{a}$
          \end{tabular}
        }
      }{
      \underset{
        \mathclap{
        \mbox{
          \tiny
          \color{greenii}
          \begin{tabular}{c}
            universal integral 4-flux
          \end{tabular}
        }
        \;\;\;\;\;
        }
      }{
        \widetilde \Gamma_4
      }
      }
    }
    \ar[d]
      |>>>>>>>{
        \phantom{ {AA} }
      }
    \ar[rr]|-{
      \overset{
        \mbox{
          \tiny
          $\phantom{a}$
        }
      }{
      \underset{
        \mathclap{
        \mbox{
          \tiny
          \color{greenii}
          \begin{tabular}{c}
            coset space realization
          \end{tabular}
        }
        }
      }{
        \simeq
      }
      }
    }
    &
    &
    \underset{
      \mathclap{
      \mbox{
        \tiny
        \color{darkblue}
        \begin{tabular}{c}
        \end{tabular}
      }
      }
    }{
      B \mathrm{Spin}(4)
    }
    \ar[ru]|-{
      \mathllap{
        \mbox{
          \tiny
          \color{greenii}
          \begin{tabular}{c}
            gravitational
            \\
            instanton density
            \\
            (\hspace{-.5pt}1\hspace{-.4pt}st \hspace{-.8pt}Pontrj. \hspace{-1pt}class)
          \end{tabular}
        }
        \!\!\!\!
      }
        \frac{1}{2}p_1
    }
    \ar[dll]
      |<<<<<<<<<<<<<<<<<<{
        \phantom{ {AA} \atop {{AA} \atop {AA}} }
      }
      |>>>>>>>>>>>>{
        \phantom{ {AA} \atop {AA} }
      }
    \ar[r]^-{
      \overset{
        \mathrlap{
          \;\;\;\;\;\;\;\;
          \overset{
            \!\!\!\!\!\!\!\!\!\!\!\!
            \mbox{
              \tiny
              \color{greenii}
              fractional
            }
          }{
          \mbox{
            \tiny
            \color{greenii}
            Euler + Pontrjagin class
          }
          }
        }
      }{
      \scalebox{.6}{
        $\tfrac{1}{2}\rchi_4 + \tfrac{1}{4}p_1$
      }
      }
    }
    &
    B^3 U(1)\,
    \mathrlap{
      \;
      {
      \overset{
        \mathclap{
        \scalebox{.5}{
          hmtpy
        }
        }
      }{
        \simeq
      }
      }
      \underset{
        \mathclap{
        \mbox{
          \tiny
          \color{darkblue}
          \begin{tabular}{c}
            classifying space for
            \\
            integral 4-cohomology
          \end{tabular}
        }
        }
      }{
        \;\; K(\mathbb{Z},4)
      }
    }
    \\
    &
    &
    \underset{
      \mathclap{
      \mbox{
        \tiny
        \color{darkblue}
        \begin{tabular}{c}
          classifying space for
          \\
          Spin connection field
        \end{tabular}
      }
      }
    }{
      B \mathrm{Spin}(8)
    }
    &
    \ar@{=>}|-{
      \underset{
        \mathclap{
        \mbox{
          \tiny
          \color{purple}
          \begin{tabular}{c}
            universal integral 3-flux
            \\
          \end{tabular}
        }
        }
      }{
          \mathclap{\phantom{\vert^{\vert^\vert}}}
          H^{{}^{\mathrm{univ}}}_3
      }
    } "s"; "t"
  }
  }
  };

 \end{tikzpicture}
 }
 }
\end{tabular}

\vspace{.18cm}

\noindent Hence we have proven the following,
for the mathematical formulation of the M-theory C-field
subject to \hyperlink{HypothesisH}{\it Hypothesis H}
according to \cite{FSS19b}, and the corresponding
formulation of the B-field in the M5-brane sigma-model according to
\cite{FSS19c}:

\vspace{-.1cm}

\begin{theorem}
  \label{GaugeFieldEmerges}
  Assuming
  C-field charge-quantization in J-twisted 4-Cohomotopy
  for M-theory on 8-manifolds,
  then the induced
  B-field charge quantization in twisted 3-Cohomotopy
  on the M5-brane worldvolume
  is equivalently charge-quantization
  in $\mathrm{String}^{c_2}(4)$-cohomology,
  according to the homotopy-commutative diagram \eqref{TheFactorization}.
\end{theorem}

\vspace{0mm}
\noindent {\bf Differential String structure on M5-branes.}
We have focused here on discussion of the topological sector of all fields,
classified by a non-abelian generalized but ``topological'' cohomology;
while the full field content is
in \emph{differential non-abelian cohomology} \cite{dcct}\cite[\S 4.2]{FSS20b}.
For example, the classifying space $B \mathrm{Spin}(8)$ in the
above discussion is to be promoted
to the smooth moduli stack $\mathbf{B}\mathrm{Spin}(8)_{\mathrm{conn}}$
of principal Spin connections \cite{FSS10}.
While we do not discuss such differential form data here, one impact of
Theorem \ref{GaugeFieldEmerges} is that it makes immediate how
this discussion should proceed: namely directly by
promoting \eqref{TheFactorization}
from a diagram in spaces to a diagram of the corresponding
smooth $\infty$-stacks according
to \cite{FSS10}\cite{FSS15} (reviewed in \cite{FSS13});
hence, in particular, promoting
the topological twisted String structure classified by the space
$B \mathrm{String}^{c_2}(4)$ \eqref{HomotopyFiberProduct}
to twisted \emph{differential} string structure classified
by a smooth 2-stack $\mathbf{B}\mathrm{String}^{c_2}_{\mathrm{conn}}$,
according to \cite{SSS12}\cite{FSS12a}\cite{FSS12b}:

\vspace{-.5cm}

\begin{equation}
\label{DifferentialRefinement}
\hspace{-.28cm}
\fbox{
\hspace{1.02cm}
\begin{tabular}{lcccc}
  &
  \begin{tabular}{c}
    {\it Classifying spaces}
  \end{tabular}
  &
  \xymatrix{
    \ar@<+3pt>@{<-}[rrr]
    ^-{
      \mbox{
        \tiny
        pass to
        \color{greenii}
        shape
      }
    }
    _-{
      \mbox{
        \tiny
        (trad.: ``{\color{greenii}geometric realization}'')
      }
    }
    &&&
  }
  &
  {\it Moduli stacks}
  \\
  &
  \begin{tabular}{l}
    $
      \mathllap{
        \raisebox{2pt}{
          \tiny
          \color{darkblue}
          \begin{tabular}{r}
            for Spin bundles
            \\
            $\mathclap{\phantom{\vert_{\vert_\vert}}}$
            Spin structure
          \end{tabular}
        }
        \,
      }
      B \mathrm{Spin}(n)
    $
    \\
    $
      \mathllap{
        \raisebox{2pt}{
          \tiny
          \color{darkblue}
          \begin{tabular}{r}
            for $B^{n-2} U(1)$-bundles
            \\
            $\mathclap{\phantom{\vert_{\vert_\vert}}}$
            (ordinary) $n$-cohomology
          \end{tabular}
        }
        \,
      }
      B^{n-1} U(1)
    $
    \\
    $
      \mathllap{
        \raisebox{2pt}{
          \tiny
          \color{darkblue}
          \begin{tabular}{r}
            for $\mathrm{String}^{c_2}$ 2-bundles
            \\
            $\mathclap{\phantom{\vert_{\vert_\vert}}}$
            twisted String structure
          \end{tabular}
        }
        \,
      }
      B \mathrm{String}^{c_2}(n)
    $
    \\
    $
      \mathllap{
        \raisebox{2pt}{
          \tiny
          \color{darkblue}
          \begin{tabular}{r}
            for $\Omega S^n$ $\infty$-bundles
            \\
            $\mathclap{\phantom{\vert_{\vert_\vert}}}$
            $n$-Cohomotopy
          \end{tabular}
        }
        \,
      }
      B \Omega S^n \;\simeq\; S^n
    $
  \end{tabular}
  &
  \raisebox{8pt}{
  $
    \overset{
      \mbox{
        \tiny
        \color{greenii}
        promote
      }
    }{
      \longmapsto
    }
  $
  }
  &
  \;\;\;\;\;
  \begin{tabular}{l}
    $
      \mathbf{B}\mathrm{Spin}(n)_{\mathrm{conn}}
      \mathrlap{
        \;\;\;\;\;\;\,\hspace{1pt}
        \raisebox{2pt}{
          \tiny
          \color{darkblue}
          $\mathclap{\phantom{\vert_{\vert_\vert}}}$
          for Spin connections
        }
      }
    $
    \\
    $
      \mathbf{B}^{n-1} U(1)_{\mathrm{conn}}
      \mathrlap{
        \;\;\;
        \raisebox{2pt}{
          \tiny
          \color{darkblue}
          \begin{tabular}{l}
            for abelian gerbe connections
            \\
            $\mathclap{\phantom{\vert_{\vert_\vert}}}$
            \phantom{for}
             differential $n$-cohomology (ordinary)
          \end{tabular}
        }
      }
    $
    \\
    $
      \mathbf{B}\mathrm{String}^{c_2}(n)_{\mathrm{conn}}
      \mathrlap{
        \!\!\!
        \raisebox{2pt}{
          \tiny
          \color{darkblue}
          \begin{tabular}{l}
            for String 2-connections
            \\
            $\mathclap{\phantom{\vert_{\vert_\vert}}}$
            \phantom{for} twisted differential String structure
          \end{tabular}
        }
      }
    $
    \\
    $
      \mathbf{B} \Omega S^n_{\mathrm{conn}}
      \mathrlap{
        \;\;\;\;\;\;\;\;\;\;\;\;\hspace{1.4pt}
        \raisebox{2pt}{
          \tiny
          \color{darkblue}
          \begin{tabular}{l}
            for $\Omega S^n$ $\infty$-connections
            \\
            \phantom{for } differential $n$-Cohomotopy
          \end{tabular}
        }
      }
    $
  \end{tabular}
\end{tabular}
\hspace{3.11cm}
}
\end{equation}

\vspace{-.1cm}

Following the method of \cite{SSS08}\cite{FSS10} for constructing
such differential refinements, the idea
of differential String structures on M5-branes
has recently been explored in \cite{SaemannSchmidt17}
in an attempt to find a higher gauge theoretic interpretation
of the action functionals for non-abelian $D = 6$, $\mathcal{N} = (1,0)$
gauge theories proposed in \cite{SSW11}.
But here the conceptual origin and precise flavor of the
string gauge field on the M5 seems to have remained open.

\medskip
Theorem \ref{GaugeFieldEmerges} solves this
issue by pinpointing specifically
$\mathrm{String}^{c_2}(4)$-structure \eqref{HomotopyFiberProduct}
and explaining how this connects to the broader structure of
the M5-brane in M-theory,
showing that
this is the charge quantization on the M5
that relates to the web of anomaly cancellation
conditions in M-theory\cite{FSS19b}\cite{FSS19c}\cite{SS20a}.
First consequences have been drawn in \cite{Roberts20}.

\medskip

With the stringy field content on the M5-brane in hand,
we close by discussing the {\it Hopf WZ-term}
in the M5-brane action functional \cite{Intriligator00}
{\it recast as a
functional on a $\mathrm{String}^{c_2}$ higher gauge field}
(noticing that the Hopf WZ term induces the full Lagrangian density,
by the method of \cite{FSS19d}).

 \vspace{.2cm}

 \hspace{-.88cm}
 \begin{tabular}{ll}

  \begin{minipage}[left]{12.5cm}

    {\bf The cocycle $c_6$ in twisted 7-Cohomotopy},
    on the classifying space $\widehat X^8$
    for M5-brane sigma-model fields, arises from
    the construction of the latter
    as a homotopy pullback
    of the Borel-equivariant quaternionic Hopf fibration.
    This $c_6$ is the {\it dual of the C-field}, with flux form $G_7$,
    in the situation that the C-field itself trivializes \cite{FSS19b},
    as it does after pullback along $\phi$ to the M5, by
    \eqref{BianchiOfH3OnSigma}.
  \end{minipage}
 &
 \hspace{-.1cm}
 \raisebox{-23pt}{
\scalebox{.8}{
\begin{tikzpicture}

 \draw[dashed, lightgray] (-5.05,3.15) ellipse (2.8 and 1.2);
 \clip (-5.05,3.15) ellipse (2.8 and 1.2);

 \draw (0,0) node
 {
  \raisebox{68pt}{
  \xymatrix@C=35pt{
    \overset{
      \!\!\!\!\!\!\!\!
      \!\!\!\!\!\!\!\!
      \!\!\!\!
      \mathclap{
      \mbox{
        \tiny
        \color{darkblue}
        \begin{tabular}{c}
          classifying space for
          \\
          M5 sigma-model fields
        \end{tabular}
      }
    }
    }{
    \mathllap{
      \mathbb{R}^{2,1}
      \times
      \,
    }
    }
    \widehat X^{8}
    \ar@{}[ddr]|-{\mbox{\tiny (pb)}}
    \ar[r]_-{
      \underset{
        \mathclap{
        \;\;\;\;\;\;\;\;\;
        \color{greenii}
        \mbox{
          \tiny
          \begin{tabular}{c}
            dual C-field in
            \\
            J-twisted 7-Cohomotopy
          \end{tabular}
        }
        }
      }
      {
        c_6
      }
    }
    \ar@[lightgray][dd]|-{
      \mathclap{
      \mbox{
        \tiny
        \color{greenii}
        \begin{tabular}{c}
          embedding
          \\
          fields
        \end{tabular}
      }
      }
    }
    &
    \overset{
      \mathrlap{
      \!\!\!\!\!\!\!\!\!\!
      \!\!\!\!\!\!\!\!\!\!
      \mbox{
        \tiny
        \color{darkblue}
        \begin{tabular}{c}
          classifying space for
          \\
          J-twisted 7-Cohomotopy
        \end{tabular}
      }
      }
    }{
      S^7 \!\sslash \mathrm{Sp}(2)
    }
    \ar@[lightgray]@{-->}[dr]|-{
      \mbox{
        \tiny
        \color{purple}
        \begin{tabular}{c}
          induced universal
          \\
          non-abelian higher gauge field
          \\
          on M5-worldvolume
        \end{tabular}
      }
    }
    \ar@[lightgray]@/^6pc/[rrrd]_<<<<<<<<<{\ }="s2"
    |<<<<<<<<<<{
      \underset{
        \mbox{
          \tiny
          \begin{tabular}{c}
            $\phantom{a}$
            \\
            $\phantom{a}$
          \end{tabular}
        }
      }{
      \overset{
        \mathclap{
        \mbox{
          \tiny
          \color{greenii}
          \begin{tabular}{c}
            universal integral shifted C-field 4-flux
            \\
            universally restricted to M5-worldvolume
          \end{tabular}
        }
        }
      }{
          (h_{\mathbb{H}})^\ast \widetilde \Gamma_4
      }
      }
    }
    \ar@[lightgray][rr]|-{
      \overset{
        \mbox{
          \tiny
          $\phantom{a}$
        }
      }{
      \underset{
        \mathclap{
        \mbox{
          \tiny
          \color{greenii}
          coset space realization
        }
        }
      }{
        \simeq
      }
      }
    }
      ^<<<<<<<<<<<<<<<<<<<<<{\ }="t2"
    \ar@[lightgray][dd]|-{
      \underset{
        \mathclap{
        \mbox{
          \tiny
          \color{greenii}
          \begin{tabular}{c}
            quaternionic
            \\
            Hopf fibration
          \end{tabular}
        }
        }
      }{
        h_{\mathbb{H}}
        \sslash
        \mathrm{Sp}(2)
      }
    }
    &&
    \overset{
      \mathllap{
      \mbox{
        \tiny
        \color{darkblue}
        \begin{tabular}{l}
          classifying space
          \\
          for gauge field on M5
        \end{tabular}
      }
      \;\;\;\;\;\;\;\;\;\;\;\;\;\;\;\;
      }
    }{
      B \mathrm{Sp}(1)
    }
    \ar[dr]|-{
      \mathllap{
        \mbox{
          \tiny
          \color{greenii}
          \begin{tabular}{c}
            $\phantom{a}$
            \\
            gauge
            \\
            instanton density
            \\
            (2nd Chern class)
          \end{tabular}
        }
        \!\!\!\!\!\!\!\!
      }
      c_2
    }
    \\
    &
    &
    \underset{
      \mathclap{
      \mbox{
        \tiny
        \color{darkblue}
        \begin{tabular}{c}
          classifying space for
          \\
          twisted String structure
        \end{tabular}
      }
      \;\;\;\;\;\;
      }
    }{
      B \mathrm{String}^{c_2}\!(4)
    }
    \ar@{}[rr]|>>>>>>>>>>>>>{\mbox{\tiny (pb)}}
    \ar[ur]_>>>>>>>>>>{\ }="s"
    \ar[dr]^-{\ }="t"
    &
    &
    \overset{
      \mathclap{
      \mbox{
        \tiny
        \color{darkblue}
        \begin{tabular}{c}
          classifying space
          \\
          for
          \\
          circle 2-gerbes
        \end{tabular}
      }
      }
    }{
      B^3 U(1)
    }
    \\
    \mathllap{
      \underset{
        \mathrlap{
          \mbox{
            \tiny
            \color{darkblue}
            \begin{tabular}{c}
              spacetime for
              \\
              M-theory on 8-mfds
            \end{tabular}
          }
        }
      }{
        \mathbb{R}^{2,1}
        \times
        \,
      }
    }
    X^8
    \ar[dr]_-{
      \underset{
        \mathclap{
        \mbox{
          \tiny
          \color{greenii}
          \begin{tabular}{c}
            tangent
            structure
          \end{tabular}
        }
        }
      }{
        T X^8
      }
    }
    \ar@/_3.95pc/[rrrr]|>>>>>>>>>>>>>>>>>>>>>>>{
      \overset{
        \mbox{
          \tiny
          $\phantom{a}$
        }
      }{
      \underset{
        \mathclap{
        \mbox{
          \tiny
          \color{greenii}
          shifted integral C-field 4-flux
        }
        }
      }{
        \widetilde G_4 \;\simeq_{{}_{\mathbb{R}}}\; G_4 + \tfrac{1}{4}p_1
      }
      }
    }
    \ar[r]^-{
      \overset{
        \mathclap{
        \;\;\;\;\;\;\;\;
        \mbox{
          \tiny
          \color{greenii}
          \begin{tabular}{c}
            C-field in
            \\
            J-twisted Cohomotopy
          \end{tabular}
        }
        }
      }{
        c
      }
    }
    &
    \underset{
      \mathrlap{
      \!\!\!\!
      \mbox{
        \tiny
        \color{darkblue}
        \begin{tabular}{c}
          classifying space
          \\
          J-twisted 4-Cohomotopy
        \end{tabular}
      }
      }
    }{
      S^4 \!\sslash \mathrm{Spin}(5)
    }
    \ar@/_1.85pc/[rrr]|>>>>>>>>>>>>>>>>>{
      \overset{
        \mbox{
          \tiny
          \begin{tabular}{c}
            $\phantom{a}$
          \end{tabular}
        }
      }{
      \underset{
        \mathclap{
        \mbox{
          \tiny
          \color{greenii}
          \begin{tabular}{c}
            universal integral 4-flux
          \end{tabular}
        }
        \;\;\;\;\;
        }
      }{
        \widetilde \Gamma_4
      }
      }
    }
    \ar[d]
      |>>>>>>>{
        \phantom{ {AA} }
      }
    \ar[rr]|-{
      \overset{
        \mbox{
          \tiny
          $\phantom{a}$
        }
      }{
      \underset{
        \mathclap{
        \mbox{
          \tiny
          \color{greenii}
          \begin{tabular}{c}
            coset space realization
          \end{tabular}
        }
        }
      }{
        \simeq
      }
      }
    }
    &
    &
    \underset{
      \mathclap{
      \mbox{
        \tiny
        \color{darkblue}
        \begin{tabular}{c}
        \end{tabular}
      }
      }
    }{
      B \mathrm{Spin}(4)
    }
    \ar[ru]|-{
      \mathllap{
        \mbox{
          \tiny
          \color{greenii}
          \begin{tabular}{c}
            gravitational
            \\
            instanton density
            \\
            (\hspace{-.5pt}1\hspace{-.4pt}st \hspace{-.8pt}Pontrj. \hspace{-1pt}class)
          \end{tabular}
        }
        \!\!\!\!\!
      }
        \tfrac{1}{2}p_1
    }
    \ar[dll]
      |<<<<<<<<<<<<<<<<{
        \phantom{ {AA} \atop {{AA} \atop {AA}} }
      }
      |>>>>>>>>>>{
        \phantom{ {AA} \atop {AA} }
      }
    \ar[r]^-{
      \overset{
        \mathrlap{
          \;\;\;\;\;\;\;\;
          \overset{
            \!\!\!\!\!\!\!\!\!\!\!\!
            \mbox{
              \tiny
              \color{greenii}
              fractional
            }
          }{
          \mbox{
            \tiny
            \color{greenii}
            Euler + Pontrjagin class
          }
          }
        }
      }{
      \scalebox{.6}{
        $\tfrac{1}{2}\rchi_4 + \tfrac{1}{4}p_1$
      }
      }
    }
    &
    B^3 U(1)
    \mathrlap{
      \;
      {
      \overset{
        \mathclap{
        \scalebox{.5}{
          hmtpy
        }
        }
      }{
        \simeq
      }
      }
      \underset{
        \mathclap{
        \mbox{
          \tiny
          \color{darkblue}
          \begin{tabular}{c}
            classfying space for
            \\
            integral 4-cohomology
          \end{tabular}
        }
        }
      }{
        \, K(\mathbb{Z},4)
      }
    }
    \\
    &
    \underset{
      \mathclap{
      \mbox{
        \tiny
        \color{darkblue}
        \begin{tabular}{c}
          classifying space for
          \\
          M2-brane background structure
        \end{tabular}
      }
      }
    }{
      B \mathrm{Spin}(8)
    }
    &
    \ar@{=>}|-{
      \scalebox{.8}{
        $
        \underset{
          \mathclap{
          \mbox{
            \tiny
            \color{purple}
            \begin{tabular}{c}
              Green-Schwarz mechanism
            \end{tabular}
          }
          }
        }{
          {H_3 =}
          \atop
          {\mathbf{CS}(\omega) - \mathbf{CS}(A)}
        }
        $
      }
    } "s"; "t"
    \ar@{=>}|-{
      \mbox{
        \tiny
        \color{purple}
        background charge
      }
    }
       "s2"; "t2"
  }
  }
 };

\end{tikzpicture}
}
}

\end{tabular}

\vspace{.1cm}

\noindent
The {\it flux of the dual C-field} on $X^8$
must measure the \emph{Page charge} \cite[(41)]{DuffStelle91} of
solitonic M2-branes in the spacetime $\mathbb{R}^{2,1} \times X^8$,
each stretched along the $\mathbb{R}^{2,1}$ factor. But this requires care:

\vspace{-1.5mm}

\hspace{-.9cm}
\begin{tabular}{ll}

\begin{minipage}[left]{12.6cm}
  By the \emph{Poincar{\'e}-Hopf theorem} for 7-Cohomotopy
  \cite[2.6, 3.7]{FSS19b},
  it follows that if the loci of these M2-branes are assumed to be removed
  from spacetime
  (as the locus of the magnetic monopole is in traditional
  Dirac charge-quantization
  \cite[16.4e]{Frankel11})
  then the Euler 8-class $\rchi_8$ of $X^8$ vanishes. This is the situation of
  M-theory on 8-manifolds
  \cite{SethiVafaWitten96}\cite[3.8 \& Rmk. 3.1]{FSS19b}.

\medskip
  \noindent {\it The vanishing of the Euler 8-class} is homotopy-theoretically imposed
  \cite[Def. 4.2]{FSS19c}
  by homotopy pullback of the whole diagram \eqref{TheFactorization}
  to the homotopy fiber $B \widehat {\mathrm{Sp}(2)}$ of the classifying map for $\chi_8$:
\end{minipage}
&
\hspace{-.2cm}
\raisebox{-60pt}{
\scalebox{.8}{
 \begin{tikzpicture}

 \draw[dashed, gray] (-3.93,4.6) ellipse (2.85 and 2.85);
 \clip (-3.93,4.6) ellipse (2.85 and 2.85);

  \draw (-7.75,-5.5) node
   {
     $
       \underset{
         \scalebox{.75}{
           \color{olive}
           \begin{tabular}{c}
             M5 brane
           \end{tabular}
         }
       }{
         \underbrace{
           \phantom{------}
         }
       }
     $
   };
  \draw (-8+2.4,-5.5) node
   {
     $
       \underset{
         \scalebox{.75}{
           \color{olive}
           \begin{tabular}{c}
             spacetime
           \end{tabular}
         }
       }{
         \underbrace{
           \phantom{------}
         }
       }
     $
   };
  \draw (-8+2.4+2.9,-5.5) node
   {
     $
       \underset{
         \scalebox{.75}{
           \color{olive}
           \begin{tabular}{c}
             $\;\;\;\;\;\;\;$C-field
           \end{tabular}
         }
       }{
         \underbrace{
           \phantom{----------}
         }
       }
     $
   };
  \draw (-8+9.1,-5.5) node
   {
     $
       \underset{
         \scalebox{.75}{
           \color{olive}
           \begin{tabular}{c}
             $\;\;\;\;\;\;\;\;\;$B-field
           \end{tabular}
         }
       }{
         \underbrace{
           \phantom{-----------}
         }
       }
     $
   };
  \draw (-8+12.55,-5.5) node
   {
     $
       \underset{
         \scalebox{.75}{
           \color{olive}
           \begin{tabular}{c}
             $\;\;\;$gauge field
           \end{tabular}
         }
       }{
         \underbrace{
           \phantom{--------}
         }
       }
     $
   };
  \draw (-8+15.25,-5.5) node
   {
     $
       \underset{
         \scalebox{.75}{
           \color{olive}
           \begin{tabular}{c}
             $\;\;\;\;\;$fluxes
           \end{tabular}
         }
       }{
         \underbrace{
           \phantom{-------}
         }
       }
     $
   };

 \draw (2.3,4.95) node
 {
   \xymatrix@C=38pt{
     \ar@[lightgray]@/_.8pc/[rrrrrr]|>>>>>>>>>>{
       \overset{
         \mbox{
           \tiny
           \phantom{W}
         }
       }{
       \underset{
         \mbox{
           \tiny
           \color{greenii}
           \begin{tabular}{c}
             universal Hopf WZ term
           \end{tabular}
         }
       }{
         \scalebox{.6}{
         $
         \widetilde \Gamma_7
         \;:=\;
         H^{{}^{\mathrm{univ}}}_3 \!
         \wedge
         (\widetilde \Gamma_4 + \tfrac{1}{2}p_1)
         +
         2\Gamma_7
         $
         }
       }
       }
     }
     &&&&&&
   }
 };

 \draw[draw=white, fill=white] (0.975,4.95) circle (.07);

  \draw (0,0) node
  {
  \raisebox{68pt}{
  \xymatrix@C=45pt{
    &
    \mathllap{
      \mathllap{
        \overset{
          \mbox{
            \tiny
            \color{darkblue}
            3-sphere fiber
          }
        }{
        \mbox{
          \tiny
          \color{darkblue}
          (over any point)
        }
        }
        \;
        {\color{lightgray}
        \mathrm{Sp}(1)_R
        \,\simeq\;
        }
      }
    }
    \;
    {\color{lightgray}S^3}
    \mathrlap{
    }
    \ar@[lightgray][d]
    \\
    &
    \mathllap{
      \overset{
        \mbox{
          \tiny
          \color{darkblue}
          classifying space for
        }
      }
      {
        \mbox{
          \tiny
          \color{darkblue}
          M5 sigma-model fields
        }
      }
      \;
      \mathbb{R}^{2,1}
      \times
      \,
    }
    \widehat X^{8}
    \ar@{}[ddr]|-{\mbox{\tiny (pb)}}
    \ar[r]_-{
      \underset{
        \mathclap{
        \;\;\;\;\;\;\;\;\;
        \color{greenii}
        \mbox{
          \tiny
          \begin{tabular}{c}
            dual C-field in
            \\
            J-twisted 7-Cohomotopy
          \end{tabular}
        }
        }
      }
      {
        c_6
      }
    }
    \ar@[lightgray][dd]|-{
      \mbox{
        \tiny
        \color{greenii}
        \begin{tabular}{c}
          3-sphere fibration
          \\
          over spacetime
        \end{tabular}
      }
    }
    &
    \overset{
      \mathrlap{
      \!\!\!\!\!\!\!\!\!\!
      \!\!\!\!\!\!\!\!\!\!
      \mbox{
        \tiny
        \color{darkblue}
        \begin{tabular}{c}
          classifying space for
          \\
          J-twisted 7-Cohomotopy
        \end{tabular}
      }
      }
    }{
      S^7 \!\sslash \mathrm{Sp}(2)
    }
    \ar@[lightgray]@/^6pc/[rrrd]_<<<<<<<<<<<<{\ }="s2"
    |<<<<<<<<<<{
      \underset{
        \mbox{
          \tiny
          \begin{tabular}{c}
            $\phantom{a}$
            \\
            $\phantom{a}$
          \end{tabular}
        }
      }{
      \overset{
        \mathclap{
        \mbox{
          \tiny
          \color{lightgray}
          \begin{tabular}{c}
            universal integral shifted C-field 4-flux
            \\
            universally restricted to M5-worldvolume
          \end{tabular}
        }
        }
      }{
          (h_{\mathbb{H}})^\ast \widetilde \Gamma_4
      }
      }
    }
    \ar@[lightgray][rr]|-{
      \overset{
        \mbox{
          \tiny
          $\phantom{a}$
        }
      }{
      \underset{
        \mathclap{
        \mbox{
          \tiny
          \color{greenii}
          coset space realization
        }
        }
      }{
        \simeq
      }
      }
    }
      ^<<<<<<<<<<<<<<<<<<<<<<{\ }="t2"
    \ar@[lightgray][dd]|-{
      \underset{
        \mathclap{
        \mbox{
          \tiny
          \color{greenii}
          \begin{tabular}{c}
            quaternionic
            \\
            Hopf fibration
          \end{tabular}
        }
        }
      }{
        h_{\mathbb{H}}
        \sslash
        \mathrm{Sp}(2)
      }
    }
    &&
    \overset{
      \mathllap{
      \mbox{
        \tiny
        \color{darkblue}
        \begin{tabular}{l}
          classifying space
          \\
          for gauge field on M5
        \end{tabular}
      }
      \;\;\;\;\;\;\;\;\;\;\;\;\;\;\;\;
      }
    }{
      B \mathrm{Sp}(1)_{L}
    }
    \ar[dr]|-{
      \mathllap{
        \mbox{
          \tiny
          \color{greenii}
          \begin{tabular}{c}
            $\phantom{a}$
            \\
            gauge
            \\
            instanton density
            \\
            (2nd Chern class)
          \end{tabular}
        }
        \!\!\!\!\!\!\!\!
      }
      c_2
    }
    \\
    \overset{
      \mathllap{
      \mbox{
        \tiny
        \color{darkblue}
        \begin{tabular}{c}
          (extended)
          \\
          M5-worldvolume
        \end{tabular}
      }
      \;\;\;\;\;\;
      }
    }{
      \widehat \Sigma
    }
    \ar@/^2.42pc/@{-->}[rrr]^>>>>>>>>>{
      \mbox{
        \tiny
        \color{purple}
        \begin{tabular}{c}
          induced
          \\
          non-abelian higher gauge field
          \\
          on M5-worldvolume
        \end{tabular}
      }
    }
    \ar[ur]|>>>>>>>>>>>{
      \mbox{
        \tiny
        \color{greenii}
        \begin{tabular}{c}
          B-field in
          \\
          twisted 3-Cohomotopy
        \end{tabular}
      }
      \;\;\;\;\;\;\;\;\;\;
    }
    ^<<<<<<<{b_2\!\!}
    \ar[dr]|>>>>>>>>>>>{
      {
      \mbox{
        \tiny
        \color{greenii}
        \begin{tabular}{c}
          embedding
          \\
          field
        \end{tabular}
      }
      \;\;\;
      }
    }
    _<<<<<<<{
      \phi \!\!\!
    }
    &
    &
    &
    \underset{
      \mathclap{
      \mbox{
        \tiny
        \color{darkblue}
        \begin{tabular}{c}
          classifying space for
          \\
          twisted String structure
        \end{tabular}
      }
      \;\;\;\;\;\;
      }
    }{
      B \mathrm{String}^{c_2}\!(4)
    }
    \ar@{}[rr]|>>>>>>>>>>>>>{\mbox{\tiny (pb)}}
    \ar[ur]_>>>>>>>>>>>{\ }="s"
    \ar[dr]^>>>>>>>>>>>{\ }="t"
    &
    &
    \overset{
      \mathclap{
      \mbox{
        \tiny
        \color{darkblue}
        \begin{tabular}{c}
          classifying space
          \\
          for
          \\
          circle 2-gerbes
        \end{tabular}
      }
      }
    }{
      B^3 U(1)
    }
    \\
    &
    \mathllap{
      \underset{
        \mathrlap{
          \mbox{
            \tiny
            \color{darkblue}
            \begin{tabular}{c}
              spacetime for
              \\
              M-theory on 8-mfds
            \end{tabular}
          }
        }
      }{
        \mathbb{R}^{2,1}
        \times
        \,
      }
    }
    X^8
    \ar[dr]_-{
      \underset{
        \mathclap{
        \mbox{
          \tiny
          \color{greenii}
          \begin{tabular}{c}
            tangent
            structure
          \end{tabular}
        }
        }
      }{
        T X^8
      }
    }
    \ar@/_3.925pc/[rrrr]|>>>>>>>>>>>>>>>>>>>>>>>{
      \overset{
        \mbox{
          \tiny
          $\phantom{a}$
        }
      }{
      \underset{
        \mathclap{
        \mbox{
          \tiny
          \color{greenii}
          shifted integral C-field 4-flux
        }
        }
      }{
        \; \widetilde G_4 \;\simeq_{{}_{\mathbb{R}}}\; G_4 + \frac{1}{4}p_1
      }
      }
    }
    \ar[r]^-{
      \overset{
        \mathclap{
        \;\;\;\;\;\;\;\;
        \mbox{
          \tiny
          \color{greenii}
          \begin{tabular}{c}
            C-field in
            \\
            J-twisted 4-Cohomotopy
          \end{tabular}
        }
        }
      }{
        c_3
      }
    }
    &
    \underset{
      \mathrlap{
      \!\!\!\!
      \mbox{
        \tiny
        \color{darkblue}
        \begin{tabular}{c}
          classifying space
          \\
          J-twisted 4-Cohomotopy
        \end{tabular}
      }
      }
    }{
      S^4 \!\sslash \mathrm{Spin}(5)
    }
    \ar@/_1.81pc/[rrr]|>>>>>>>>>>>>>>>>>{
      \overset{
        \mbox{
          \tiny
          \begin{tabular}{c}
            $\phantom{a}$
          \end{tabular}
        }
      }{
      \underset{
        \mathclap{
        \mbox{
          \tiny
          \color{greenii}
          \begin{tabular}{c}
            universal integral 4-flux
          \end{tabular}
        }
        \;\;\;\;\;
        }
      }{
        \widetilde \Gamma_4
      }
      }
    }
    \ar[d]
      |>>>>>>>{
        \phantom{ {AA} }
      }
    \ar[rr]|-{
      \overset{
        \mbox{
          \tiny
          $\phantom{a}$
        }
      }{
      \underset{
        \mathclap{
        \mbox{
          \tiny
          \color{greenii}
          \begin{tabular}{c}
            coset space realization
          \end{tabular}
        }
        }
      }{
        \simeq
      }
      }
    }
    &
    &
    \underset{
      \mathclap{
      \mbox{
        \tiny
        \color{darkblue}
        \begin{tabular}{c}
        \end{tabular}
      }
      }
    }{
      B \mathrm{Spin}(4)
    }
    \ar[ru]|-{
      \mathllap{
        \mbox{
          \tiny
          \color{greenii}
          \begin{tabular}{c}
            gravitational
            \\
            instanton density
            \\
            (\hspace{-.5pt}1\hspace{-.4pt}st \hspace{-.8pt}Pontrj. \hspace{-1pt}class)
          \end{tabular}
        }
        \!\!\!\!
      }
        \frac{1}{2}p_1
    }
    \ar[dll]
      |<<<<<<<<<<<<<<<<<<{
        \phantom{ {AA} \atop {{AA} \atop {AA}} }
      }
      |>>>>>>>>>>>>{
        \phantom{ {AA} \atop {AA} }
      }
    \ar[r]^-{
      \overset{
        \mathrlap{
          \;\;\;\;\;\;\;\;
          \overset{
            \!\!\!\!\!\!\!\!\!\!\!\!
            \mbox{
              \tiny
              \color{greenii}
              fractional
            }
          }{
          \mbox{
            \tiny
            \color{greenii}
            Euler + Pontrjagin class
          }
          }
        }
      }{
      \scalebox{.6}{
        $\tfrac{1}{2}\rchi_4 + \tfrac{1}{4}p_1$
      }
      }
    }
    &
    B^3 U(1)
    \\
    &
    &
    \underset{
      \mathclap{
      \mbox{
        \tiny
        \color{darkblue}
        \begin{tabular}{c}
          classifying space for
          \\
          Spin connection field
        \end{tabular}
      }
      }
    }{
      B \mathrm{Spin}(8)
    }
    &
    \ar@{=>}|-{
      \scalebox{.8}{
        $
        \underset{
          \mathclap{
          \mbox{
            \tiny
            \color{purple}
            \begin{tabular}{c}
              universal integral 3-flux
              \\
            \end{tabular}
          }
          }
        }{
          \mathclap{\phantom{\vert^{\vert^\vert}}}
          H^{{}^{\mathrm{univ}}}_3
        }
        $
      }
    } "s"; "t"
  }
  }
  };

 \draw (7.4,-3.05) node
 {
      \begin{tabular}{c}
        \begin{rotate}{-90}
          $\!\!\!\mathclap{\simeq}$
        \end{rotate}
        \\
        $
        \underset{
          \mbox{
            \tiny
            \color{darkblue}
            \begin{tabular}{c}
              classifying space for
              \\
              integral 4-cohomology
            \end{tabular}
          }
        }{
          \;\; K(\mathbb{Z},4)
        }
        $
      \end{tabular}
 };

 \draw (-2,5.1) node
 {
   \xymatrix@R=28pt@C=37pt{
     \overset{
       \mathclap{
       \mbox{
         \tiny
         \color{darkblue}
         \begin{tabular}{c}
           classifying space for
           \\
           J-twisted 7-cohomotopy
           \\
           with vanishing Euler 8-class
         \end{tabular}
       }
       }
     }{
       S^7 \!\sslash \widehat{\mathrm{Sp}(2)}
     }
     \ar[d]
     \ar@[lightgray][r]
     &
     \\
     &
   }
 };

 \draw (1.5,3.5) node
 {
   \xymatrix@R=25pt{
     \overset{
       \mathclap{
       \mbox{
         \tiny
         \color{darkblue}
         \begin{tabular}{c}
           classifying space for
           \\
           twisted String structure
           \\
           with vanishing Euler 8-class
         \end{tabular}
       }
       }
     }{
       B \mathrm{String}^{c_2}_{\rchi_8}\!(4)
     }
     \ar[dddd]
      |<<<<<<<<{
        \phantom{
          {A \atop } \atop { A \atop {A \atop A} }
        }
      }
      |>>>>>>>>>>>>>>>>>>>>>>{
        \phantom{{}_{\vert_{\vert_{\vert_{\vert_{\vert_{\vert}}}}}}}
      }
     &
     \\
     \\
     \\
     \\
     &
   }
 };

 \draw (8.3,4.9) node
 {
   \xymatrix@R=40pt@C=37pt{
     \overset{
       \mathclap{
       \mbox{
         \tiny
         \color{darkblue}
         \begin{tabular}{c}
           classifying space
           \\
           for
           \\
           circle 6-gerbes
         \end{tabular}
       }
       }
     }{
       B^7 U(1)
     }
     \ar@{}[d]
     &
     \\
     &
   }
 };

 \draw (-4.47,4.15) node
 {
   \xymatrix@C=46pt@R=54pt{
     &
     \\
     \ar@{..>}[ur]
   }
 };

 \draw[draw=white, fill=white] (-4.47,4.15) circle (.32);

 \draw (-4.47,4.15) node
 {
   \xymatrix@C=46pt@R=54pt{
     &
     \\
     \ar@{}[ur]
       |-{
       \mathclap{
       \;\;\;\;\;\;\;
       \mbox{
         \tiny
         \color{greenii}
         \begin{tabular}{c}
           Euler 8-class vanishes
           \\
           (M2-brane loci removed)
         \end{tabular}
       }
       }
     }
   }
 };

 \draw (4.45,4.8) node
 {
   \xymatrix@R=40pt@C=64pt{
     \ar@{}[d]
     \ar[rr]^-{
       \mbox{
         \tiny
         \color{purple}
         \begin{tabular}{c}
           stringy
           Hopf WZ-term of M5-brane
         \end{tabular}
       }
     }
     &&
     \\
     &
   }
 };

 \end{tikzpicture}
}
}

\end{tabular}

\vspace{-14pt}

\begin{equation}
\label{PullbackToHomotopyFiber}
\raisebox{-0pt}{
\hspace{-.66cm}
\begin{tabular}{ll}

\hspace{-1.7cm}
\begin{minipage}[left]{5cm}

  Thus in the case of vanishing Euler 8-class of $X^8$,
  the cocycle $c_6$ in J-twisted 7-Cohomotopy
  on the classifying space for M5-brane sigma model fields
  lifts to the dotted map shown above.

\end{minipage}

  &
  \hspace{.2cm}
  $$
    \raisebox{31pt}{
    \xymatrix{
      \overset{
        \mathclap{
        \mbox{
          \tiny
          \color{darkblue}
          \begin{tabular}{c}
            classifying space for
            \\
            J-twisted 7-Cohomotopy
            \\
            with vanishing Euler 8-class
          \end{tabular}
        }
        }
      }{
        \big( S^7 \!\sslash \widehat{\mathrm{Sp}(2)}\big)
      }
      \ar@{}[drr]|-{ \mbox{\tiny (pb)} }
      \ar[d]
      \ar[rr]
      &&
      \overset{
        \mathclap{
        \mbox{
          \tiny
          \color{darkblue}
          \begin{tabular}{c}
            classifying space for
            \\
            J-twisted 7-Cohomotopy
            \\
            \phantom{a}
          \end{tabular}
        }
        }
      }{
        S^7 \!\sslash \mathrm{Sp}(2)
      }
      \ar[d]
      \\
      B \widehat{ \mathrm{Sp}(2) }
      \ar[rr]_-{
        \mbox{
          \tiny
          \color{greenii}
          \begin{tabular}{c}
            homotopy fiber of
            \\
            universal Euler 8-class
          \end{tabular}
        }
      }
      &&
      B \mathrm{Spin}(8)
      \ar[rr]^-{ \rchi_8 }_-{ \mbox{\tiny\color{greenii} universal Euler 8-class} }
      &&
      K(\mathbb{Z},8)
    }
    }
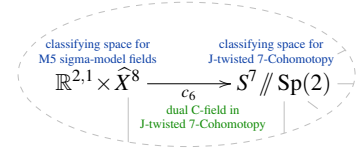
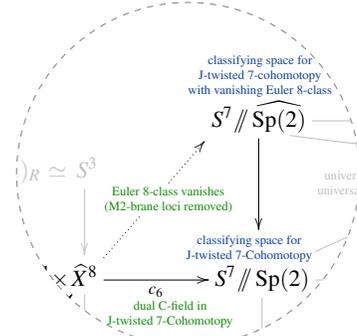
  $$

\end{tabular}
}
\end{equation}

\vspace{-.2cm}

\noindent {\bf The universal Hopf WZ-term of the M5-brane}
is, assuming \hyperlink{HypothesisH}{\it Hypothesis H},
a class $\widetilde \Gamma_7$ in integral 7-cohomology
on this space \eqref{PullbackToHomotopyFiber},
as shown in green below. This is the result of \cite[Theorem 4.8]{FSS19c}:

\vspace{-.0cm}

\hspace{-.8cm}
\begin{tabular}{ll}

 \hspace{-.2cm}
 \begin{minipage}[left]{4.7cm}
Equivalently, this $\widetilde \Gamma_7$ is the
universal {\it Page charge} density
sourced by M2-branes,
and as such is the ``dual'' of the integral shifted 4-flux
$\widetilde \Gamma_4$.
The integrality of $\widetilde \Gamma_7$
is crucial both for the Hopf WZ term to be well defined as a
Wess-Zumino term for the M5-brane, as well as its interpretation
as M2-brane charge being consistent.

\medskip
Now, the dashed factorization in \eqref{TheFactorization} directly implies,
under pullback to $B \widehat{\mathrm{Sp}(2)}$ \eqref{PullbackToHomotopyFiber} that:
\end{minipage}
&
\;\;\;\;
\raisebox{-100pt}{
 \scalebox{.7}{
 \begin{tikzpicture}

 \draw (2.3,4.95) node
 {
   \xymatrix@C=38pt{
     \ar@/_.8pc/[rrrrrr]|>>>>>>>>>>{
       \overset{
         \mbox{
           \tiny
           \begin{tabular}{c}
             \phantom{a}
             \\
             \phantom{a}
           \end{tabular}
         }
       }{
       \underset{
         \mbox{
           \tiny
           \begin{tabular}{c}
             \color{greenii}
             universal integral 7-flux
             \\
             ({\color{greenii}Page charge}, {\color{greenii}Hopf WZ term})
           \end{tabular}
         }
       }{
         \scalebox{.6}{
         $
         \widetilde \Gamma_7
         \;:=\;
         H^{{}^{\mathrm{univ}}}_3 \!
         \wedge
         (\widetilde \Gamma_4 + \tfrac{1}{2}p_1)
         +
         2\Gamma_7
         $
         }
       }
       }
     }
     &&&&&&
   }
 };

 \draw[draw=white, fill=white] (0.975,4.95) circle (.07);

  \draw (0,.45) node
  {
  \raisebox{68pt}{
  \xymatrix@C=45pt{
    &
    \mathllap{
      \mathllap{
        \overset{
          \mbox{
            \tiny
            \color{lightgray}
            3-sphere fiber
          }
        }{
        \mbox{
          \tiny
          \color{lightgray}
          (over any point)
        }
        }
        \;
        {\color{lightgray}
        \mathrm{Sp}(1)_R
        \,\simeq\;}
      }
    }
    \;
    {\color{lightgray}S^3}
    \mathrlap{
    }
    \ar@[lightgray][d]
    &&&&
    \\
    &
    \mathllap{
      \overset{
        \mbox{
          \tiny
          \color{darkblue}
          classifying space for
        }
      }
      {
        \mbox{
          \tiny
          \color{darkblue}
          M5 sigma-model fields
        }
      }
      \;
      \mathbb{R}^{2,1}
      \times
      \,
    }
    \widehat X^{8}
    \ar@[lightgray]@{}[ddr]|-{\mbox{\tiny \color{lightgray}(pb)}}
    \ar@[lightgray][r]_-{
      \underset{
        \mathclap{
        \;\;\;\;\;\;\;\;\;
        \color{lightgray}
        \mbox{
          \tiny
          \begin{tabular}{c}
            dual C-field in
            \\
            J-twisted 7-Cohomotopy
          \end{tabular}
        }
        }
      }
      {
        {\color{lightgray}c_6}
      }
    }
    \ar[dd]|-{
      \mbox{
        \tiny
        \color{lightgray}
        \begin{tabular}{c}
          3-sphere fibration
          \\
          over spacetime
        \end{tabular}
      }
    }
    &
    \overset{
      \mathrlap{
      \!\!\!\!\!\!\!\!\!\!
      \!\!\!\!\!\!\!\!\!\!
      \mbox{
        \tiny
        \color{lightgray}
        \begin{tabular}{c}
          classifying space for
          \\
          J-twisted 7-Cohomotopy
        \end{tabular}
      }
      }
    }{
      {\color{lightgray}S^7 \!\sslash \mathrm{Sp}(2)}
    }
    \ar@[lightgray]@/^6pc/[rrrd]_<<<<<<<<<<<<{\ }="s2"
    |<<<<<<<<<<{
      \underset{
        \mbox{
          \tiny
          \begin{tabular}{c}
            $\phantom{a}$
            \\
            $\phantom{a}$
          \end{tabular}
        }
      }{
      \overset{
        \mathclap{
        \mbox{
          \tiny
          \color{lightgray}
          \begin{tabular}{c}
            universal integral shifted C-field 4-flux
            \\
            universally restricted to M5-worldvolume
          \end{tabular}
        }
        }
      }{
        {\color{lightgray}(h_{\mathbb{H}})^\ast \widetilde \Gamma_4}
      }
      }
    }
    \ar@[lightgray][rr]|-{
      \overset{
        \mbox{
          \tiny
          $\phantom{a}$
        }
      }{
      \underset{
        \mathclap{
        \mbox{
          \tiny
          \color{lightgray}
          coset space realization
        }
        }
      }{
        {\color{lightgray}\simeq}
      }
      }
    }
      ^<<<<<<<<<<<<<<<<<<<<<<{\ }="t2"
    \ar@[lightgray][dd]|-{
      \underset{
        \mathclap{
        \mbox{
          \tiny
          \color{lightgray}
          \begin{tabular}{c}
            quaternionic
            \\
            Hopf fibration
          \end{tabular}
        }
        }
      }{
        {\color{lightgray}h_{\mathbb{H}}
        \sslash
        \mathrm{Sp}(2)}
      }
    }
    &&
    \overset{
      \mathllap{
      \mbox{
        \tiny
        \color{lightgray}
        \begin{tabular}{l}
          classifying space
          \\
          for gauge field on M5
        \end{tabular}
      }
      \;\;\;\;\;\;\;\;\;\;\;\;\;\;\;\;
      }
    }{
      {\color{lightgray}B \mathrm{Sp}(1)_{L}}
    }
    \ar@[lightgray][dr]|-{
      \mathllap{
        \mbox{
          \tiny
          \color{lightgray}
          \begin{tabular}{c}
            $\phantom{a}$
            \\
            gauge
            \\
            instanton density
            \\
            (2nd Chern class)
          \end{tabular}
        }
        \!\!\!\!\!\!\!\!
      }
      {\color{lightgray}c_2}
    }
    \\
    \overset{
      \mathllap{
      \mbox{
        \tiny
        \color{darkblue}
        \begin{tabular}{c}
          (extended)
          \\
          M5-worldvolume
        \end{tabular}
      }
      \;\;\;\;\;\;
      }
    }{
      {\widehat \Sigma}
    }
    \ar@/_3.4pc/@{-->}[rrruu]|>>>>>>>>>>>>>>>>>>>>>>>{
      \mbox{
        \tiny
        \color{purple}
        \begin{tabular}{c}
          induced
          \\
          non-abelian higher gauge field
          \\
          on M5-worldvolume
        \end{tabular}
      }
    }
    \ar[ur]|>>>>>>>>>>>{
      \mbox{
        \tiny
        \color{black}
        \begin{tabular}{c}
          B-field in
          \\
          twisted 3-Cohomotopy
        \end{tabular}
      }
      \;\;\;\;\;\;\;\;\;\;
    }
    ^<<<<<<<{b_2\!\!}
    \ar[dr]|>>>>>>>>>>>{
      {
      \mbox{
        \tiny
        \color{black}
        \begin{tabular}{c}
          embedding
          \\
          field
        \end{tabular}
      }
      \;\;\;
      }
    }
    _<<<<<<<{
      \phi \!\!\!
    }
    &
    &
    &
    \underset{
      \mathclap{
      \mbox{
        \tiny
        \color{lightgray}
        \begin{tabular}{c}
          classifying space for
          \\
          twisted String structure
        \end{tabular}
      }
      \;\;\;\;\;\;
      }
    }{
      {\color{lightgray}B \mathrm{String}^{c_2}\!(4)}
    }
    \ar@[lightgray]@{}[rr]|>>>>>>>>>>>>>{\mbox{\tiny \color{lightgray}(pb)}}
    \ar@[lightgray][ur]_>>>>>>>>>>>{\ }="s"
    \ar@[lightgray][dr]^>>>>>>>>>>>{\ }="t"
    &
    &
    \overset{
      \mathclap{
      \mbox{
        \tiny
        \color{lightgray}
        \begin{tabular}{c}
          classifying space
          \\
          for
          \\
          circle 2-gerbes
        \end{tabular}
      }
      }
    }{
      {\color{lightgray}B^3 U(1)}
    }
    \\
    &
    \mathllap{
      \underset{
        \mathrlap{
          \mbox{
            \tiny
            \color{darkblue}
            \begin{tabular}{c}
              spacetime for
              \\
              M-theory on 8-mfds
            \end{tabular}
          }
        }
      }{
        \mathbb{R}^{2,1}
        \times
        \,
      }
    }
    X^8
    \ar@[lightgray][dr]_-{
      \underset{
        \mathclap{
        \mbox{
          \tiny
          \color{lightgray}
          \begin{tabular}{c}
            tangent
            structure
          \end{tabular}
        }
        }
      }{
        {\color{lightgray}T X^8}
      }
    }
    \ar@[lightgray]@/_3.925pc/[rrrr]|>>>>>>>>>>>>>>>>>>>>>>>{
      \overset{
        \mbox{
          \tiny
          $\phantom{a}$
        }
      }{
      \underset{
        \mathclap{
        \mbox{
          \tiny
          \color{lightgray}
          shifted integral C-field 4-flux
        }
        }
      }{
        {\color{lightgray}
        \; \widetilde G_4 \;\simeq_{{}_{\mathbb{R}}}\; G_4 + \frac{1}{4}p_1}
      }
      }
    }
    \ar@[lightgray][r]^-{
      \overset{
        \mathclap{
        \;\;\;\;\;\;\;\;
        \mbox{
          \tiny
          \color{lightgray}
          \begin{tabular}{c}
            C-field in
            \\
            J-twisted 4-Cohomotopy
          \end{tabular}
        }
        }
      }{
        {\color{lightgray}c_3}
      }
    }
    &
    \underset{
      \mathrlap{
      \!\!\!\!
      \mbox{
        \tiny
        \color{lightgray}
        \begin{tabular}{c}
          classifying space
          \\
          J-twisted 4-Cohomotopy
        \end{tabular}
      }
      }
    }{
      {\color{lightgray}S^4 \!\sslash \mathrm{Spin}(5)}
    }
    \ar@[lightgray]@/_1.81pc/[rrr]|>>>>>>>>>>>>>>>>>{
      \overset{
        \mbox{
          \tiny
          \begin{tabular}{c}
            $\phantom{a}$
          \end{tabular}
        }
      }{
      \underset{
        \mathclap{
        \mbox{
          \tiny
          \color{lightgray}
          \begin{tabular}{c}
            universal integral 4-flux
          \end{tabular}
        }
        \;\;\;\;\;
        }
      }{
        {\color{lightgray}\widetilde \Gamma_4}
      }
      }
    }
    \ar@[lightgray][d]
      |>>>>>>>{
        \phantom{ {AA} }
      }
    \ar@[lightgray][rr]|-{
      \overset{
        \mbox{
          \tiny
          $\phantom{a}$
        }
      }{
      \underset{
        \mathclap{
        \mbox{
          \tiny
          \color{lightgray}
          \begin{tabular}{c}
            coset space realization
          \end{tabular}
        }
        }
      }{
        {\color{lightgray}\simeq}
      }
      }
    }
    &
    &
    \underset{
      \mathclap{
      \mbox{
        \tiny
        \color{lightgray}
        \begin{tabular}{c}
        \end{tabular}
      }
      }
    }{
      {\color{lightgray}B \mathrm{Spin}(4)}
    }
    \ar@[lightgray][ru]|-{
      \mathllap{
        \mbox{
          \tiny
          \color{lightgray}
          \begin{tabular}{c}
            gravitational
            \\
            instanton density
            \\
            (\hspace{-.5pt}1\hspace{-.4pt}st \hspace{-.8pt}Pontrj. \hspace{-1pt}class)
          \end{tabular}
        }
        \!\!\!\!
      }
        {\color{lightgray}\frac{1}{2}p_1}
    }
    \ar@[lightgray][dll]
      |<<<<<<<<<<<<<<<<<<{
        \phantom{ {AA} \atop {{AA} \atop {AA}} }
      }
      |>>>>>>>>>>>>{
        \phantom{ {AA} \atop {AA} }
      }
    \ar@[lightgray][r]^-{
      \overset{
        \mathrlap{
          \;\;\;\;\;\;\;\;
          \overset{
            \!\!\!\!\!\!\!\!\!\!\!\!
            \mbox{
              \tiny
              \color{lightgray}
              fractional
            }
          }{
          \mbox{
            \tiny
            \color{lightgray}
            Euler + Pontrjagin class
          }
          }
        }
      }{
      \scalebox{.6}{
        {\color{lightgray}$\tfrac{1}{2}\rchi_4 + \tfrac{1}{4}p_1$}
      }
      }
    }
    &
    {\color{lightgray}B^3 U(1)}
    \\
    &
    &
      {\color{lightgray}B \mathrm{Spin}(8)}
    &
    \ar@[lightgray]@{=>}|-{
      \scalebox{.8}{
        $
        \underset{
          \mathclap{
          \mbox{
            \tiny
            \color{lightgray}
            \begin{tabular}{c}
              universal integral 3-flux
              \\
            \end{tabular}
          }
          }
        }{
          {\color{lightgray}\mathclap{\phantom{\vert^{\vert^\vert}}}
          H^{{}^{\mathrm{univ}}}_3}
        }
        $
      }
    } "s"; "t"
  }
  }
  };

 \draw (7.4,-3.05) node
 {
      \begin{tabular}{c}
        \begin{rotate}{-90}
          {\color{lightgray}$\!\!\!\mathclap{\simeq}$}
        \end{rotate}
        \\
        $
        \underset{
          \mbox{
            \tiny
            \color{lightgray}
            \begin{tabular}{c}
              classifying space for
              \\
              integral 4-cohomology
            \end{tabular}
          }
        }{
          {\color{lightgray}\;\; K(\mathbb{Z},4)}
        }
        $
      \end{tabular}
 };

 \draw (-2,5.1) node
 {
   \xymatrix@R=28pt@C=37pt{
     \overset{
       \mathclap{
       \mbox{
         \tiny
         \color{darkblue}
         \begin{tabular}{c}
           classifying space for
           \\
           J-twisted 7-cohomotopy
           \\
           with vanishing Euler 8-class
         \end{tabular}
       }
       }
     }{
       S^7 \!\sslash \widehat{\mathrm{Sp}(2)}
     }
     \ar[d]
     \ar@{-->}[r]
     &
     \\
     &
   }
 };

 \draw (1.5,3.5) node
 {
   \xymatrix@R=25pt{
     \overset{
       \mathclap{
       \mbox{
         \tiny
         \color{darkblue}
         \begin{tabular}{c}
           classifying space for
           \\
           twisted String structure
           \\
           with vanishing Euler 8-class
         \end{tabular}
       }
       }
     }{
       B \mathrm{String}^{c_2}_{\rchi_8}\!(4)
     }
     \ar[dddd]
      |<<<<<<<<{
        \phantom{
          {A \atop } \atop { A \atop {A \atop A} }
        }
      }
      |>>>>>>>>>>>>>>>>>>>>>>{
        \phantom{{}_{\vert_{\vert_{\vert_{\vert_{\vert_{\vert}}}}}}}
      }
     &
     \\
     \\
     \\
     \\
     &
   }
 };

 \draw (8.3,4.9) node
 {
   \xymatrix@R=40pt@C=37pt{
     \overset{
       \mathclap{
       \mbox{
         \tiny
         \color{darkblue}
         \begin{tabular}{c}
           classifying space
           \\
           for
           \\
           circle 6-gerbes
         \end{tabular}
       }
       }
     }{
       B^7 U(1)
     }
     \ar@{}[d]
     &
     \\
     &
   }
 };

 \draw (-4.47,4.15) node
 {
   \xymatrix@C=46pt@R=54pt{
     &
     \\
     \ar@{..>}[ur]
   }
 };

 \draw[draw=white, fill=white] (-4.47,4.15) circle (.32);

 \draw (-4.47,4.15) node
 {
   \xymatrix@C=46pt@R=54pt{
     &
     \\
     \ar@{}[ur]
       |-{
       \mathclap{
       \;\;\;\;\;\;\;
       \mbox{
         \tiny
         \color{black}
         \begin{tabular}{c}
           Euler 8-class vanishes
           \\
           (M2-brane loci removed)
         \end{tabular}
       }
       }
     }
   }
 };

 \draw (4.45,4.8) node
 {
   \xymatrix@R=40pt@C=64pt{
     \ar@{}[d]
     \ar[rr]^-{
       \mbox{
         \tiny
         \color{purple}
         \begin{tabular}{c}
           stringy
           Hopf WZ-term of M5-brane
         \end{tabular}
       }
     }
     &&
     \\
     &
   }
 };

 \end{tikzpicture}
 }
 }
 \end{tabular}

\medskip
\noindent {\bf The Hopf WZ term descends to a class
on $\mathrm{String}^{c_2}_{\rchi_8}$ higher gauge fields}
as shown in purple in the above.

\newpage

\noindent {\bf Acknowledgement.}
This project started at the Jan. 2020
 workshop {\it M-theory and Mathematics} supported by the NYUAD Research Institute.
D. Fiorenza thanks New York University Abu Dhabi for hospitality
during that time.

\vspace{.1cm}


\medskip

\vspace{1cm}
\noindent Domenico Fiorenza, {\it Dipartimento di Matematica, La Sapienza Universita di Roma, Piazzale Aldo Moro 2, 00185 Rome, Italy.}
\\
{\tt fiorenza@mat.uniroma1.it}
\\
\\
\noindent  Hisham Sati, {\it Mathematics, Division of Science, New York University Abu Dhabi, UAE.}
\\
{\tt hsati@nyu.edu}
\\
\\
\noindent  Urs Schreiber, {\it Mathematics, Division of Science, New York University Abu Dhabi, UAE;
on leave from Czech Academy of Science, Prague.}
\\
{\tt us13@nyu.edu}

\end{document}